\documentclass{article}
\usepackage[margin=1in]{geometry}
\usepackage{graphicx}
\usepackage[
backend=biber,
]{biblatex}
\usepackage{booktabs}
\usepackage{tabularray}
\usepackage{url}
\usepackage{hyperref}
\usepackage{tikz} 
\usepackage{float}
\usepackage{authblk}
\usepackage{amsmath}
\usepackage{adjustbox}
\usepackage{xcolor}
\usepackage{appendix}

\title{A comparison of stretched-grid and limited-area modelling for data-driven regional weather forecasting}
\author[*,1]{Jasper S. Wijnands}
\author[*,2]{Michiel Van Ginderachter}
\author[*,1]{Bastien Fran\c{c}ois}
\author[1]{Sophie Buurman}
\author[2,3]{Piet Termonia}
\author[*,2]{Dieter Van den Bleeken}
\affil[1]{Royal Netherlands Meteorological Institute (KNMI), De Bilt, the Netherlands}
\affil[2]{Royal Meteorological Institute of Belgium, Uccle, Belgium}
\affil[3]{Ghent University, Ghent, Belgium}
\affil[*]{Equal contributions}

\date{\today}

\begin{document}

\maketitle

\begin{abstract}
Regional machine learning weather prediction (MLWP) models based on graph neural networks have recently demonstrated remarkable predictive accuracy, outperforming numerical weather prediction models at lower computational costs. In particular, limited-area model (LAM) and stretched-grid model (SGM) approaches have emerged for generating high-resolution regional forecasts, based on initial conditions from a regional (re)analysis. While LAM uses lateral boundaries from an external global model, SGM incorporates a global domain at lower resolution. This study aims to understand how the differences in model design impact relative performance and potential applications. Specifically, the strengths and weaknesses of these two approaches are identified for generating deterministic regional forecasts over Europe. Using the Anemoi framework, models of both types are built by minimally adapting a shared architecture and trained using global and regional reanalyses in a near-identical setup. Several inference experiments have been conducted to explore their relative performance and highlight key differences. Results show that both LAM and SGM are competitive deterministic MLWP models with generally accurate and comparable forecasting performance over the regional domain. Various differences were identified in the performance of the models across applications. LAM is able to successfully exploit high-quality boundary forcings to make predictions within the regional domain and is suitable in contexts where global data is difficult to acquire. SGM is fully self-contained for easier operationalisation, can take advantage of more training data and significantly surpasses LAM in terms of (temporal) generalisability. Our paper can serve as a starting point for meteorological institutes to guide their choice between LAM and SGM in developing an operational data-driven forecasting system.  
\end{abstract}

\section{Introduction}
In recent years the domain of weather forecasting has been subject to a paradigmatic transformation due to the arrival of machine learning weather prediction (MLWP) models. The integration of machine learning (ML) into nowcasting applications has a history dating back to the late 1990s \cite{mccann1992, marzban1996}. In recent years, ML techniques have been increasingly employed to emulate or replace specific physical parameterisations within numerical weather prediction (NWP) models \cite[e.g.,][]{gentine2018, seifert2022}. The development of ML-based models capable of directly providing forecasts, thus replacing NWP models, nonetheless only gained traction following the pioneering contributions of \cite{dueben2018}, \cite{weyn2019}, and \cite{rasp2021}, who demonstrated the feasibility of using neural network architectures for this purpose. Subsequently, a breakthrough was achieved with the emergence and use of more specialised architectures, notably graph neural networks (GNNs) \cite{keisler2022} and neural operators \cite{pathak2022}. This sparked a surge of activity, by research institutes, operational centres as well as technology companies, to develop MLWP models based on a diverse set of deep learning architectures. Models like GraphCast \cite{lam2023} and the Artificial Intelligence Forecasting System (AIFS) \cite{lang2024} use GNNs, while Pangu-Weather \cite{bi2023} and FourCastNet \cite{pathak2022} are based on Vision Transformers. The FuXi \cite{chen2023} and Feng-Wu \cite{chen2023b} models, in turn, use other transformer-type designs.

This latest generation of MLWP models all provide global deterministic weather forecasts that outperform current state-of-the-art physics-based numerical weather prediction (NWP) models in not only the standard verification scores, but also for a selection of synoptic-scale extreme events \cite{bouallegue2024}. The success of the aforementioned models is partly a result of the availability of open and reliable global weather datasets spanning multiple decades, such as the European Centre for Medium-Range Weather Forecasts (ECMWF) fifth generation reanalysis (ERA5) \cite{ERA5}. In fact, all the previously mentioned models use some version of ERA5, resulting in an initial focus on providing global forecasts at horizontal resolutions around 0.25$^\circ$ or approximately 31 km. While these MLWP models outperform physics-based NWP models for medium-range forecasts of synoptic-scale weather patterns, a resolution of 31 km is too coarse to properly represent deep convection, one of the main drivers for the most extreme precipitation events. In order to capture deep convection a substantial increase in resolution beyond the mesoscale is necessary. Operational NWP models currently run at kilometer-scale resolutions.

In principle, there is no inherent obstacle preventing MLWP models from utilizing high-resolution datasets during the training phase. However, currently such consistent high-resolution global datasets, spanning several decades, do not exist. To overcome this limitation, methodologies have been developed involving training models with enhanced datasets that include not only ERA5 reanalysis but also higher-resolution data from operational analyses and forecasts \cite{han2024, bodnar2025}. High-resolution reanalysis datasets do exist, but are limited to regional domains (e.g. Copernicus European Regional Reanalyis (CERRA) \cite{cerra}, Copernicus Arctic Regional Reanalysis (CARRA) \cite{carra1}, Austrian Regional Reanalysis (ARA) \cite{ara}). These regional (re)analyses datasets are typically created using classical NWP limited-area models (LAMs), used by many national weather services to create high-resolution forecasts for specific regions of interest \cite{oskarsson2023, Bush2024, Mller2017, Seity2011}. To leverage these regional datasets containing additional information on finer spatio-temporal scales, two MLWP approaches have emerged providing regional high-resolution forecasts: LAMs and stretched-grid models (SGMs).

Graph-based ML LAMs \cite{oskarsson2023, adamov2025} -- inspired by NWP LAMs -- leverage the flexibility of GNNs by encoding both high-resolution data on a regional domain as well as lower-resolution lateral boundary forcing inputs from an external global dataset. During training and inference, ML LAMs generate forecasts exclusively on the regional domain. To continue the forecast to later lead times, these regional forecasts of the model are complemented by boundary forcing inputs after each time step. Two choices of ML LAM graph architectures were presented by \cite{oskarsson2023}: multi-scale mesh graphs and hierarchical mesh graphs. Multi-scale mesh graphs combine the different processor mesh levels, while hierarchical mesh graphs introduce an additional layer for each processor mesh level to capture the different scales \cite{oskarsson2023, adamov2025}. In this study, only multi-scale mesh graph-based ML LAMs are considered.

A second approach able of providing regional high-resolution MLWP forecasts is stretched-grid modelling \cite{nipen2024}. SGM differs from LAM in that the forecast is provided on both a global domain and a regional domain by a single model. Furthermore, the processor mesh is refined to a higher resolution over the regional domain compared to the global domain, creating a seamless transition between the regional and global domains. Note that in this paper, the global domain is defined as the entire globe excluding the regional domain.

Both LAM and SGM have shown to outperform NWP models on mean squared error (MSE) based losses against surface synoptic (SYNOP) observations for key variables like temperature at 2 meter above the surface and wind speed at 10 meter height \cite{nipen2024, adamov2025}. It should be noted that both models suffer from smoothing and underestimation of extremes, a well-known limitation of deterministic MLWP trained with an MSE loss function \cite{xu2024, subich2025}. However, a direct and fair comparison between LAM and SGM for regional MLWP has not yet been reported in the literature. This may be due to the independent code development of the methods. For example, several LAM approaches emerged directly from an adaptation of the GraphCast code \cite{lam2023, oskarsson2023}, while the first SGM was built on the AIFS codebase \cite{lang2024, nipen2024}. The independent development of LAM and SGM has resulted in differences in the datasets, model configuration, training procedure and verification, complicating the comparison. Hence, for new MLWP model developments, there is a lack of insight on the advantages and disadvantages of both approaches. 

The objective of this study is to evaluate the LAM and SGM approaches for regional MLWP. By focusing on their differences in applications and performance, this study aims to highlight their relative strengths and weaknesses within a fair comparison framework. In our study, the LAM and SGM approaches are developed and compared using the Anemoi framework (\url{https://anemoi.ecmwf.int/}). Anemoi is a collaborative, open-source initiative comprising several interacting Python packages addressing different aspects of the MLWP pipeline. It is being developed through a collaboration of ECMWF and meteorological institutes across Europe \cite{ECMWF2024b, Nemesio2025}. Since the LAM and SGM approaches have many overlapping structural features, Anemoi is an appropriate platform for comparing the inherent properties of these models and evaluating their relative effectiveness in meteorological applications. The models are evaluated on a set of metrics providing an extensive and impartial representation of the performance, given a common experimental configuration in Anemoi. Practically, this experimental configuration consists of common training data, model and training configurations that are as identical as possible, and a common evaluation framework. Differences in model and training configurations that are inherent to each approach will be specified in detail in the next section. Overall, we will assess performance on both reanalysis and operational analysis datasets at a single time step as well as for later lead times, together providing a comprehensive assessment of comparative performance during training, validation and inference on test data.

\section{Methods}
\label{sec:methods}
\subsection{Data}\label{subsec:data}
The models used in this study were trained using both global and regional datasets. The ERA5 global reanalysis, developed by ECMWF, is one of the most prominent reanalysis products in Europe\cite{ERA5}. It is based on the Integrated Forecasting System (IFS) and benefits from over a decade of advancements in model physics, core dynamics, and data assimilation techniques. ERA5 offers hourly outputs with a horizontal resolution of 31 km and 137 vertical levels throughout the data set. The ERA5 output fields are complemented by uncertainty estimates derived from a 10-member ensemble (available at 3-hourly intervals and reduced spatial resolution). The reanalysis spans the period from 1940 up to present day. Because of its high quality, ERA5 became the main reference data set for global MLWP. Rather than using ERA5 on its native N320 reduced Gaussian grid, in this study an O96 octahedral reduced Gaussian grid was used. This corresponds with a horizontal resolution around 100 km, effectively reducing the number of gridpoints by 93\%. This allowed for faster experimentation as the main focus in this work is the regional domain.

\begin{figure}[H]
\begin{center}
  \includegraphics[clip, width=0.5\textwidth]{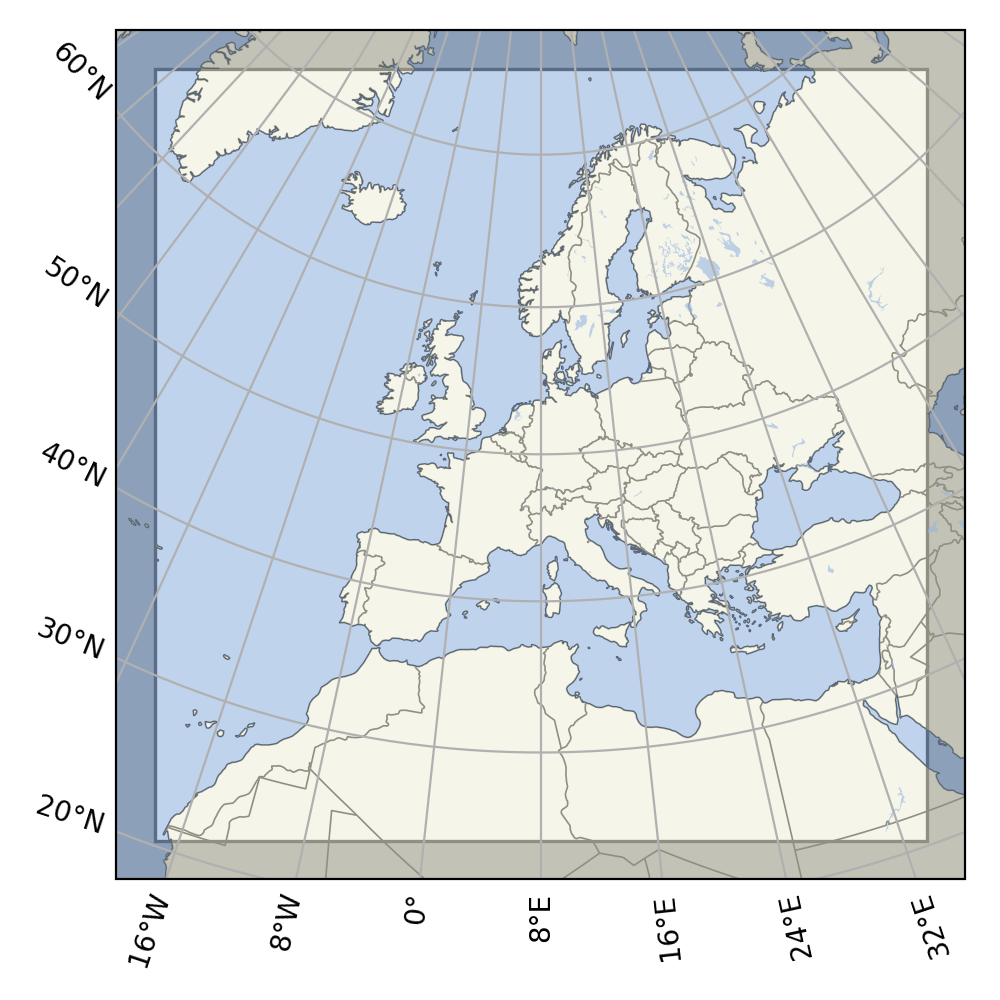}
  \caption{Presentation of the domain covered by the CERRA reanalysis (unshaded area), the regional domain in this study.} \label{fig:CERRA-domain}
\end{center}
\end{figure}

The regional-domain data is provided by the Copernicus European Regional ReAnalysis (CERRA), which has a resolution of 5.5 km \cite{cerra}. This reanalysis is based on the HARMONIE NWP system \cite{HARMONIE}, using a model configuration with a physics package of the ALADIN system \cite{ALADIN}. It provides 3-hourly data output and spans a period of 36 years and 4 months from September 1984 until December 2020 on a regional domain covering Europe, Southern Greenland and Northern Africa (see Fig.~\ref{fig:CERRA-domain}). CERRA has demonstrated an overall improvement with respect to ERA5, in particular for 2m temperature and wind, both at 10m and in the free atmosphere. Although CERRA performs slightly worse for relative humidity on the overall domain, it slightly improves on ERA5 over complex terrain, such as the Alps. This is consistent with the fact that relative humidity exhibits significant spatial variability and is another indication that CERRA better captures the small scales in comparison \cite{cerra}. An overview of the dataset specifications of ERA5 and CERRA can be found in Table~\ref{tab:datasets}.

\begin{table}[ht]
\caption{Overview of the datasets used for training.}
\label{tab:datasets}
    \centering
    \begin{adjustbox}{width=0.8\textwidth}
    \begin{tabular}{lll}
    \toprule   
    {} & ERA5 & CERRA \\
    \midrule
    Spatial coverage & Global & Europe \\
    Grid type & Octahedral reduced Gaussian (O96) & Lambert conformal conic \\
    Resolution & $\sim$100km & 5.5km \\
    \# Gridpoints & 40,320 & 1,142,761 \\
    Temporal frequency & hourly & 3-hourly \\
    Available period & 1979 - 2022 & Sep. 1984 - 2020 \\
    \bottomrule
    \end{tabular}
\end{adjustbox}
\end{table}

All models in this paper use five prognostic upper air variables (temperature, u and v components of winds, specific humidity and geopotential height) defined on 12 pressure levels (see list in Table~\ref{tab:variables}) and seven prognostic surface variables (skin/surface and 2m temperature, 2m dewpoint temperature, 10m u and v components of winds, surface and mean sea level pressure) as input and output. Two dataset specific variables (land-sea mask and surface geopotential) together with nine variables defining location, time and solar angle are used as additional input. 

Special care was needed for upper air specific humidity, 2m dewpoint temperature, 10m wind components and surface geopotential height as these variables were not directly available from the original CERRA dataset. These were computed using a (combination of) available fields (see the ``Derived'' column in Table~\ref{tab:variables}).

The input data for the operational-like inference (see Section~\ref{subsubsec:operational}) was constructed as follows: the initialisation data for the global (SGM), boundary (LAM) and regional domain (both) was taken from ECMWF's High Resolution Integrated Forecast System (IFS-HRES) \cite{IFScite} analysis and was bilinearly interpolated from its native octahedral reduced Gaussian grid (O1280, $\sim$9 km resolution) to the O96-grid and CERRA 5.5 km resolution grid for global/boundary and regional domain respectively. The boundary conditions for the LAM-inference were provided by interpolating IFS-forecasts (O1280) to the O96-grid.

Finally, verification against SYNOP observations was performed using 1706 SYNOP stations covering Europe's mainland, the United Kingdom, Iceland, the South-West coast of Greenland and Madeira. As this study focusses on the difference between SGM and LAM rather than absolute performance of the models, no height correction between gridpoints and SYNOP-stations height was performed for 2m temperature.  

\subsection{Models}
\label{subsec:models}

All models considered in this paper use a GNN architecture similar to the one introduced in GraphCast \cite{lam2023}. Specifically, using the terminology of \cite{lang2024}, information is transferred from the (input) data grid to an auxiliary lower resolution grid, the hidden (processor) grid, then processed and finally transferred back to the original data grid, via an encoder--processor--decoder architecture. First, the encoder projects data points in the data grid onto the hidden grid, extracting information from nearby grid points. Next, a processor distributes the extracted information across the whole domain using both short-range and long-range connections, corresponding to the evolution of atmospheric dynamics during a single time step. Finally, a decoder maps the obtained latent space representation back onto the data grid. To investigate model complexity, different model sizes are tested. Our initial experiments use 512 channels for the encoder, the 16-layer processor, and the decoder, resulting in 62 million trainable parameters. Larger models with 1024 channels are also investigated, resulting in 246 million trainable parameters. The model is able to distinguish between regional and global domains based on edge length, such that patterns and dependencies can be modelled on both small and large scales. Rather than a pure message-passing GNN, our models use a graph transformer architecture, with 16 heads, that incorporates attention mechanisms to learn which node and edge information is most important \cite{shi2021masked}.

In all experiments, both the LAM and SGM use the CERRA grid at native 5.5 km resolution as the regional domain. Outside the regional domain, ERA5 on a reduced Gaussian grid with O96 resolution is used as the data source (see Section~\ref{subsec:data}). The key difference between SGM and LAM is the architecture of the graphs that connect these data points into a GNN with encoder, processor and decoder modules as described above.

\begin{figure}[H]
  \includegraphics[trim = 2mm 2mm 2mm 33mm, clip, width=\textwidth]{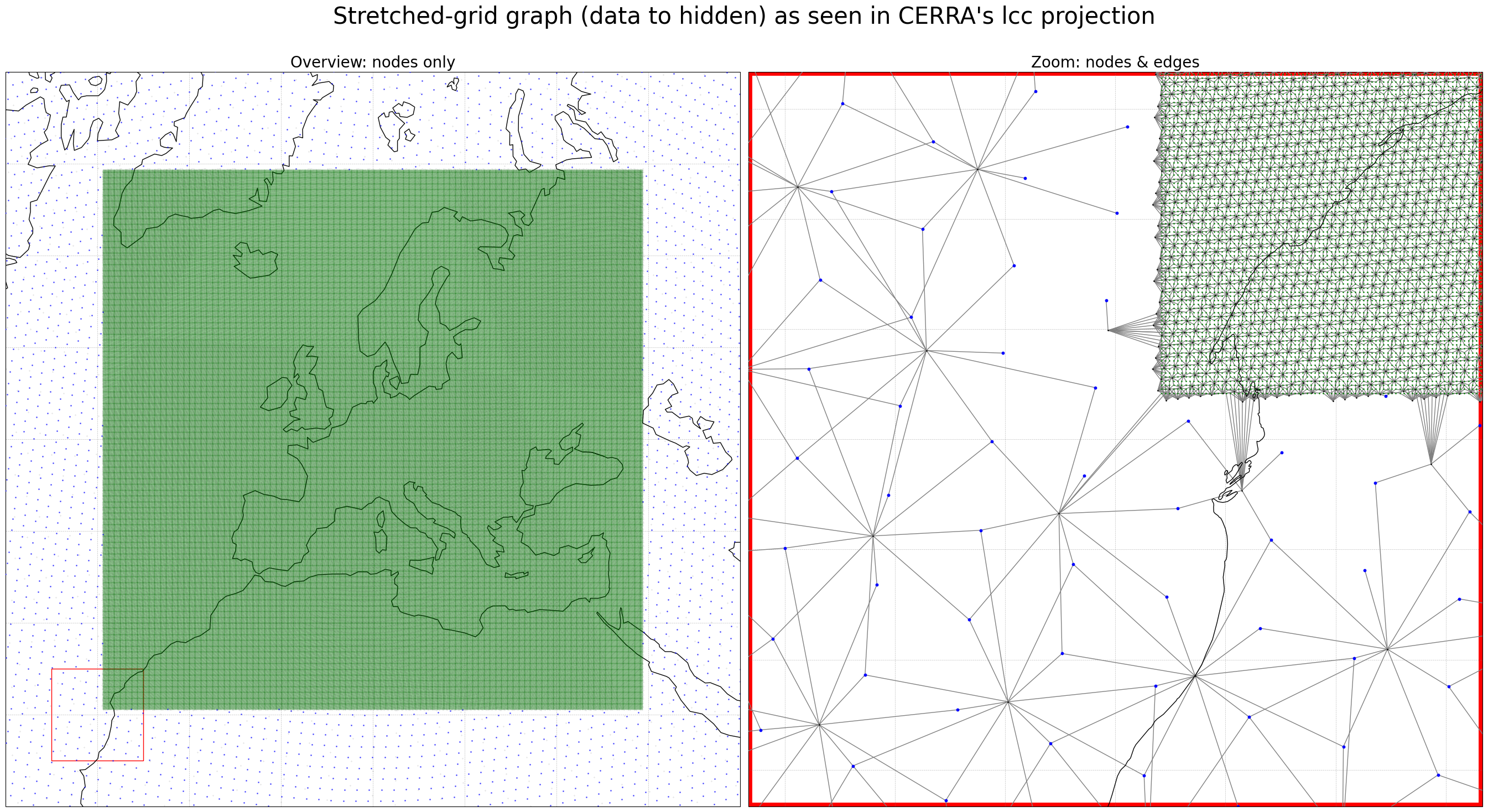}
  \caption{Illustration of the encoder of part of the SGM graph. The left panel shows the nodes of the data grid inside (green) and outside (blue) the regional domain. The right panel shows a zoomed subsection at the transition region, including the edges connecting data points to the hidden grid nodes (grey).} \label{fig:sgm_graph}
\end{figure}

The SGM graph (Fig.~\ref{fig:sgm_graph}) is composed of nodes in a high-resolution regional domain and a lower resolution global domain that are mutually exclusive (similar to \cite{nipen2024}). Together, the nodes in these domains form a grid that spans the whole globe. The auxiliary hidden grid has a similar mixed resolution. Specifically, it is a global icosahedral grid with mesh refinement nine (i.e., approximately 15 km between nodes) over an 11 km wide extension of the regional domain and mesh refinement five (approximately 240 km between nodes) over the remainder of the globe. The nodes in these data and hidden grids are connected based on the $k$-nearest neighbours algorithm, which defines the edges of the encoder and decoder. For the encoder, each node in the hidden grid extracts information from the 12 nearest nodes in the data grid. Then, the processor distributes this information along a multi-mesh edge structure, created by connecting the data nodes hierarchically among refinement levels, as in \cite{lam2023}. In the decoder, each data point only receives information from the three nearest nodes in the hidden grid. This edge connection scheme applies to the full grid, including both regional and global domains.

\begin{figure}[H]
  \includegraphics[trim = 2mm 2mm 2mm 31mm, clip, width=\textwidth]{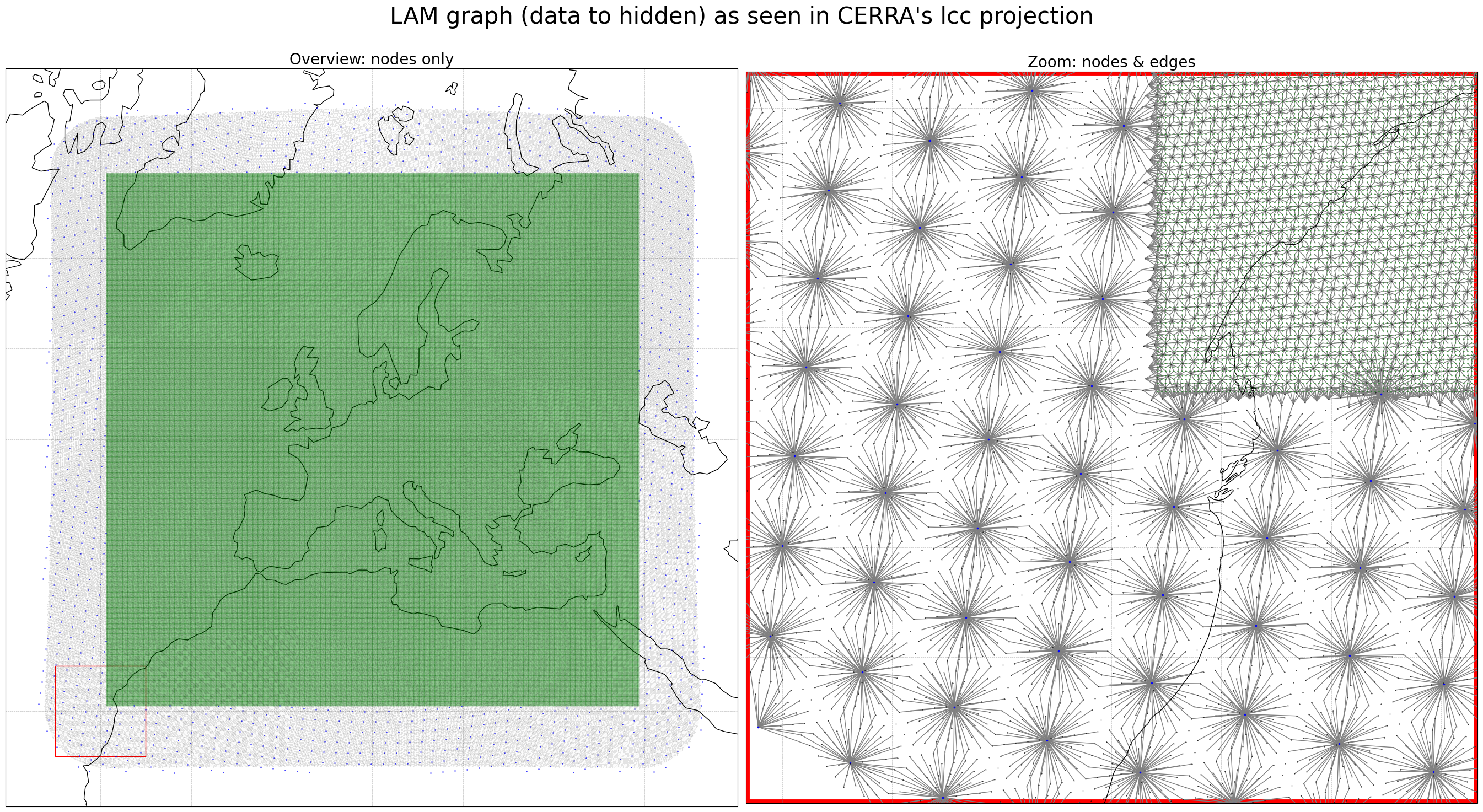}
  \caption{As Fig.~\ref{fig:sgm_graph}, but for the LAM graph.}
  \label{fig:lam_graph}
\end{figure}

The LAM graph (Fig.~\ref{fig:lam_graph}), on the contrary, has data and hidden nodes covering only a limited part of the globe. The data nodes in the high-resolution regional domain are identical to the SGM graph. However, compared to the SGM data grid, the global domain is replaced by a relatively small boundary domain. This boundary domain consists of an O96 reduced Gaussian grid surrounding the regional domain up to a width of 666 km. The hidden grid is icosahedral as in the SGM graph, but has a fixed mesh refinement of nine everywhere and extends 600 km outside the regional domain. Hence, the hidden grid over the boundary domain has a higher resolution than the corresponding region in the SGM graph. In terms of edge connections, the configuration is similar to SGM and has only two differences. First, outside of the regional domain, each hidden node is connected to all data points within a 66 km radius, in this way also defining the extent of the boundary domain. Second, the decoder data points in the boundary domain are not connected to any hidden nodes, since forecasts are only used on the regional domain. Note that the boundary size and edge connection pattern have a large impact on model performance and the choices made above are the result of testing various alternatives, as further discussed in Section~\ref{sec:discussion}. Both SGM and LAM graphs were created using the anemoi-graphs package.

\subsection{Training}
\label{subsec:training}
The model training was configured to be as similar as possible, with minor differences to account for specific characteristics of LAM and SGM. The common elements include the use of the dataset described in Section~\ref{subsec:data} with a temporal frequency of 6 hours, dividing it into training (1984--2018), validation (2019) and test (2020) sets. For the purpose of training and validating with a common 6-hourly time step, only times of day 00, 06, 12 and 18 hours UTC were included. Model weights were randomly initialised in all experiments to prevent that a pre-trained starting point benefits LAM and SGM differently. This approach contrasts with the SGM of \cite{nipen2024}, where training on the regional dataset started from weights obtained in a pre-training task for a lower resolution global model. The training of the models was split into two phases. First, the models were trained towards forecasting a single 6h time step, based on the two most recent time steps as input. In the second phase, they were further fine-tuned towards autoregressively forecasting multiple time steps, the so-called rollout training \cite{keisler2022}. For all models and training phases a variable-weighted MSE loss function was used, adopting the same variable weighting as in AIFS \cite{lang2024}. Other commonalities included a cosine learning rate schedule with a warm-up period of 1000 or 10 steps for the first and second training phase, respectively.

To assess the impact of model size, small and large-sized versions of each model type were trained using the training regime described above. Considering the first training phase, the small-sized configurations with 512 channels used a learning rate of $1\times 10^{-3}$ in the AdamW optimiser \cite{loshchilov2019decoupled}. These models were trained using 150,000 training steps with an effective batch size of four, meaning the model processed a total of 600,000 samples during training. The 1024-channel models aimed to further increase forecast performance. Therefore, an effective batch size of 16 was used during 260,000 training steps (i.e., 4.16 million samples in total). The learning rate in these experiments was doubled to $2\times10^{-3}$, as it was found that a slightly higher starting learning rate led to better forecast skill. All experiments used model sharding across multiple GPUs on a single HPC node to take advantage of multi-GPU speed-up and have access to the combined GPU memory on one node to fit a single batch (consisting of only one sample). The reported effective batch sizes were achieved by data parallelisation across multiple nodes.

During rollout training, model output was fed back into the model to forecast the next time step. In this training phase, models were trained to optimize the average loss of up to 12 rollout steps, corresponding to a lead time of 72 h. The model checkpoint from the first training phase was used to initialise the trainable parameters. Initially, the number of rollout steps was set to two, which was periodically increased until a maximum of 12 rollout steps was reached. These increments occurred every 454 batches, so a total number of 5000 iterations was used in this second training phase.

Besides differences in graph architecture (see Section~\ref{subsec:models}), there are some differences in the training process of LAM and SGM. A key, and essentially defining, difference between LAM and SGM is the rollout procedure. The SGM generates forecasts on the full global and regional domain, which coincides with its input grid. Therefore, the model can directly be rolled out autoregressively. However, since the decoder edges of the LAM are only connected to the regional domain, its output needs to be complemented with the state of the boundary domain from an external source. The merged data can then be used as input for a forecast at the next time step. This procedure of replacing the fields on the boundary domain with that of an external model is referred to as boundary forcing and can be seen as coupling the LAM to an external boundary model. In the LAM rollout training, the boundary forcing is provided by the O96 ERA5 reanalysis dataset also used for the initial states.  

An additional distinction in the training process is the evaluation and definition of the loss functions. As indicated previously, LAM forecasts for data points in the boundary region are excluded from the loss. For SGM, the loss function consists of a global and a regional component, where the weighting of these components can be adjusted using an additional hyperparameter. Experimentation showed that a 25\% regional and 75\% global weighting provided appropriate results (not shown here), which has been adopted for all experiments in this paper. A more detailed discussion about domain weighting is provided in Section~\ref{sec:discussion}. Further, new functionality in anemoi-training was developed for a fair evaluation of LAM and SGM losses. Specifically, new validation metrics were developed for SGM to split the loss into components for the regional and global domain, allowing for a comparison with the LAM during the training process.

Finally, data pre-processing followed a different procedure for LAM and SGM. Per-variable normalisation for LAM used the mean and standard deviation of the respective variable in the regional dataset, except for geopotential which was normalised based on the maximum value. Insolation, land-sea mask, time and location variables were not normalised (see Table~\ref{tab:variables}). SGM followed the same normalisation procedures but used the corresponding statistics of the global dataset. The reason for this modelling choice was that the trained SGM can be used as a starting point for transfer learning, requiring consistency in terms of data normalisation (see Section~\ref{sec:discussion}). For LAM, data normalisation based on ERA5 statistics was an unnatural choice, since data points almost exclusively originate from the regional dataset. During inference, forecasts were converted back to their original distributions, enabling a direct comparison between LAM and SGM.

\subsection{Inference}
\label{subsec:inference}

While monitoring training and validation performance through the variable-weighted MSE loss is essential during the training phase, it does not provide a suitable evaluation of the ML models' performance on unseen data. Following standard machine learning practices, inference of the trained models is performed throughout the test period (year 2020) to forecast the different weather variables listed in Table~\ref{tab:variables} and evaluate the models. The models are initialised for inference using data from two distinct time steps: the initialisation time and previous time step, 6 hours earlier. Then, forecasts of 6-hourly frequency are produced autoregressively over the regional domain for lead times +6h up to +72h. All forecasts cover the period from January 2, 2020 to December 25, 2020. Selecting these dates offers the advantage of maintaining a consistent number of lead times for the various runs, as well as ensuring that all forecasted times fall strictly within the test period. In this study, various inference experiments -- with differences in how forecasts are initialised and/or forced -- are performed to assess the capabilities of the trained ML models.

\subsubsection{Inference based on reanalysis data}\label{subsubsec:ideal}

The first inference experiment, hereafter designated as \texttt{ideal}, consists of utilizing the test period of both ERA5 and CERRA reanalysis datasets as initial conditions. By using data sources that are consistent between training and testing, this experiment aims to provide an initial performance assessment. For both the SGM and LAM, inference is initialised from CERRA on the regional domain and ERA5 on the global or boundary domain, respectively. As explained in Section~\ref{subsec:training}, the key difference between a SGM and a LAM resides in how the rollout is performed: while the SGMs autoregressively generate forecasts based on its own outputs, LAMs require boundary forcings from an external (global) dataset/model in addition to its own output on the regional domain to perform recurrent inference. In this first inference experiment, the LAM is supplied with ERA5 data for the boundary domain, enabling it to predict meteorological variables at various lead times. Note that although this initial experiment allows us to assess methods using consistent data between training and testing, this setup is not entirely realistic, since the reanalysis initial data is not available in real time. Hence, it cannot be implemented in an operational context. For LAM this also applies to the boundary forcings. It can be expected, although maybe not a priori guaranteed, that in this ``idealised'' experiment the LAM has an unrealistic advantage over the SGM. After each time step, it has access to boundary forcings including assimilated observations, while the SGM has to rely on its own forecasts. 

For this inference experiment, forecasts are initialised at 00 and 12 UTC. This setup aligns with the operational timing of meteorological institutes like ECMWF. It also ensures that the 6-hourly inferred data, at times of day 00, 06, 12, and 18 UTC, is consistent with that used during the training phase, enabling a comprehensive evaluation of performance. 

\subsubsection{Operational settings using IFS analyses and forecasts}\label{subsubsec:operational}

To evaluate the SGM and LAM in a more realistic operational setting, a second inference experiment is defined, hereafter referred to as \texttt{operational}. It consists of simply replacing the ERA5 and CERRA reanalysis datasets used in the inference experiment described in subsection~\ref{subsubsec:ideal} with operational data from IFS-HRES. For this inference experiment, the inference is initialised at 00 UTC. For both the SGM and LAM, the initial conditions are now provided by the IFS-HRES analysis, by reducing its resolution to O96 for the global/boundary domain and interpolated to the CERRA grid for the regional domain, as explained in subsection~\ref{subsec:data}. For subsequent lead times, the LAM is provided with IFS-HRES forecasts reduced to O96 resolution as boundary forcing. By utilizing IFS-HRES analyses and forecasts, this inference experiment allows us to evaluate SGM and LAM within an operational-like framework. It offers insights valuable to meteorological institutes by assessing the models under conditions that closely mimic their everyday operations. However, please note that apart from a degradation of initial conditions (i.e., interpolated analysis rather than native reanalysis), this inference experiment potentially introduces an additional error due to the use of a different type of data than what the model was trained on. Further, an additional effect would appear for the LAM trained with rollout on ERA5 forcing (see Section~\ref{subsec:training}), since then the boundary forcings would differ from the ones used during training. To better investigate the influence of the initialization error on its own, the \texttt{operational} experiments are therefore performed with the models without rollout training.

\subsubsection{Influence of boundary forcings on LAM performance}

By construction, LAM forecasts rely on boundary forcings by an external dataset or model. The choice of forcings will influence the comparison with a SGM, which provides its own boundary forecasts. To disentangle the effect of the boundary forcings from the inherent forecasting quality of the LAM within the regional domain, we perform two additional inference experiments, exploring different types of LAM boundary forcings. Both these forecasts are initialised from reanalysis, as in subsection~\ref{subsubsec:ideal}. The first option of boundary forcings in rollout to later lead times, named \texttt{pragmatic}, consists of using IFS-HRES forecasts. Please note that it is distinct from the inference experiment described in subsection~\ref{subsubsec:operational} in that it employs ERA5/CERRA reanalyses as initial conditions rather than IFS-HRES analyses, thereby avoiding the introduction of errors at initialisation. The second option, hereafter referred to as \texttt{mixed}, consists of using the SGM's global domain forecasts at the boundary region, obtained during the inference experiment described in subsection~\ref{subsubsec:ideal}. By providing boundary forcings from SGM to LAM, this second option explicitly neutralizes the possible difference in quality of boundary forcings between LAM and SGM, allowing for a more equitable evaluation of their predictions over the regional domain at later lead times. It provides deeper insights into the performance of both models specifically within the regional domain, independently of differences in the quality of global and boundary domains. In both the \texttt{pragmatic} and \texttt{mixed} inference setup, inference is initialised at 00 UTC. Furthermore, to keep a fair comparison and exclude a bias towards ERA5 boundary forcings introduced in the second training phase (see Section~\ref{subsec:training}), only the models without rollout training are used in the experiments of this subsection.

\subsubsection{Temporal generalisability}\label{sec:generalisability}

Certain aspects of the generalisability of the models beyond the particulars of the training dataset are explored in a final inference experiment. During training, the CERRA dataset was restricted to 6-hourly data at times of day 00, 06, 12 and 18 UTC; hereafter referred to as ``reference times of day''. The full CERRA and ERA5 reanalyses, however, contain data at 3-hourly and hourly frequency, respectively. This allows an experiment, hereafter designated as \texttt{extended}, where forecasts are initialised (in the test period only) at time 03, 09, 15 and 21 UTC; hereafter referred to as ``shifted times of day''. The models are then evaluated after a single 6h forecast step at times of day 09, 15, 21 and 03 UTC. Since these initial and forecasted times are not present in the training dataset they can be considered unseen by the model. This experiment thus provides insights into the temporal generalisability of SGM and LAM. All forecasts are initialised using ERA5 and CERRA reanalyses on both times of day present, respectively absent, in the training dataset. Scores are then computed for reference and shifted times of day separately. This experiment only considers forecasts of a single 6h time step, so models without rollout training are used. In addition, a separate 512-channel LAM is trained using ``shifted times of day'' only. This additional model is used exclusively to complement and strengthen the results for this inference experiment.

\section{Evaluation}
\label{sec:evaluation}
 
This section starts with an analysis of the training process. Then, model performance during the 2020 test period is evaluated and compared against CERRA reanalyses, SYNOP observations within the CERRA domain, and climatological data using various verification metrics. This assessment involves utilizing a multitude of experiments to assess and contrast LAM and SGM across various training, model and inference settings. Table~\ref{tab:experiments} provides an overview of these different setups. To easily refer to a specific type of experiment, the following naming convention has been adopted in this paper: \texttt{[Structure]-[Size]-[Rollout]-[Setup]}. For example, \texttt{LAM-1024-R01-ideal} refers to a LAM structure with 1024 channels, trained on a one time step forecasting task (i.e., results of the first training phase), where both initialisation and boundary forcing is performed with reanalysis data. In the remainder, parts of this naming convention could be omitted if they are clear from context. Please note that the exploratory nature of this work implied examining the results for a large number of combinations of the different experiment settings. For the sake of conciseness, this paper only presents explicit results for a relevant selection of these settings. Detailed information on the CERRA climatology and verification metrics used can be found in Appendix~\ref{app:metrics}.

\begin{table}[h]
  \caption{Overview of experiment settings.}
  \label{tab:experiments}
  \makebox[\textwidth][c]{
    \begin{tabular}{l c}
      \toprule
      Experiment type & Settings \\
      \midrule
      Model structure & LAM, SGM \\
      Model size (channels) & 512, 1024\\
      Autoregressive training setup & R01 (no rollout), R12 (rollout)\\
      Inference setup & ideal, operational, pragmatic, mixed, extended \\
      \bottomrule
    \end{tabular}
  }
\end{table}

\subsection{Training process}
\label{subsec:trainingprocess}
Training was logged and diagnosed on a common server using MLflow experiment tracking \cite{chen2020developments}. This formed a first evaluation stage where, through monitoring and comparison, new model settings and bottlenecks regarding computational performance were assessed.

\begin{figure}[H]
  \includegraphics[trim = 2mm 2mm 2mm 2mm, clip, width=\textwidth]{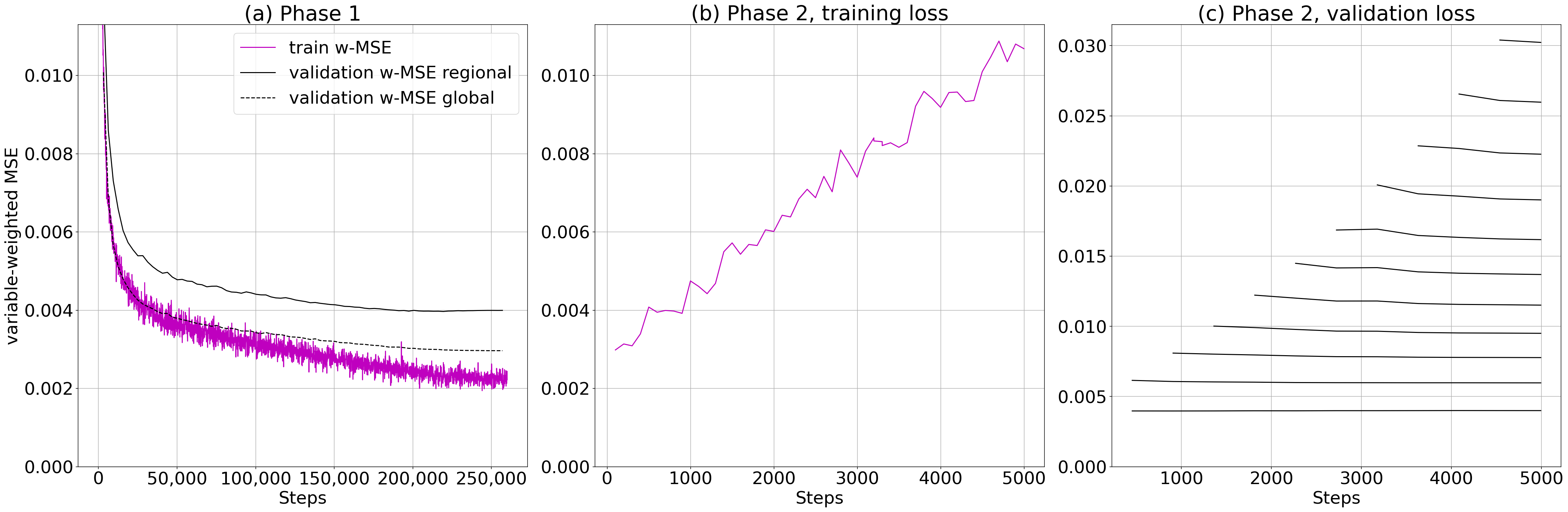}
  \caption{Illustration of the training process of \texttt{SGM-1024}. Panel (a) shows the training weighted-MSE (purple) and validation weighted-MSE in the regional (black, solid) and global domains (black, dashed) of training phase 1 (\texttt{SGM-1024-R01}). Panels (b, c) show training phase 2 (\texttt{SGM-1024-R12}), with (b) illustrating the step-up in training loss when gradually increasing the number of rollout steps, and (c) the improvement in validation weighted-MSE averaged over one to 11 time steps (bottom to top).}
  \label{fig:training_process}
\end{figure}

Model training involved two different training phases and Fig.~\ref{fig:training_process} illustrates each of them through an example. Fig.~\ref{fig:training_process}a shows the convergence of training and validation losses for the \texttt{SGM-1024-R01} training run, noting that the learning rate also reduces towards zero at the end of the training process. In comparison to similar plots for LAM (results not shown), small relative differences were typically observed in the final weighted-MSE, which will be further examined in Section~\ref{sec:overall_performance} through inference experiments. Although the gap between training and validation losses widened during training, the overall training and validation loss of \texttt{SGM-1024-R01} strictly decreased throughout the training process. For \texttt{LAM-1024-R01}, there was a minor increase in overall validation loss towards the very end of training phase 1 (not shown). For both LAM and SGM, some individual variables experienced a slight deterioration of validation performance towards the end of the training process, typically with a corresponding improvement for other variables or for the global domain (SGM only). Further, for some variables such as wind speed, the initial convergence speed was faster for SGM than for LAM. However, both converged to approximately the same level (not shown).

In Fig.~\ref{fig:training_process}b, jumps can be observed in training weighted-MSE when the number of rollout steps increases. At these points, the loss is suddenly averaged over an additional time step, which is further into the future and therefore has a higher associated loss. When monitoring the validation loss at a fixed number of time steps (Fig.~\ref{fig:training_process}c), it can be observed that the model converges as expected during rollout training. Compared to the results of the first training phase, a marginal performance deterioration at the first time step typically enables better forecasting at later lead times.

Besides performance related to the training and validation loss, the computational requirements of \texttt{LAM-1024} and \texttt{SGM-1024} were compared. Although the extensive training runs of the various models were performed on multiple HPC infrastructures, dedicated partial training runs were performed on a single HPC system to obtain a fair estimate of the computational performance. \texttt{LAM-1024} and \texttt{SGM-1024} have identical model size in terms of number of parameters, requiring approximately 1 GB storage for 246 million parameters. However, the models have a different memory footprint during training. For example, it was observed that \texttt{LAM-1024-R01} uses 25\% less system memory than \texttt{SGM-1024-R01} during the first training phase. This is partly caused by a smaller set of data nodes, requiring less data loading when preparing batches. At the same time, the contrary was observed for GPU memory, with the SGM requiring 17\% less memory. The reason is that the LAM graph has an extended high-resolution hidden mesh, leading to a greater number of processor nodes and edges compared to the SGM graph. This also explains the 10\% reduction in overall training speed observed for \texttt{LAM-1024-R01} compared to \texttt{SGM-1024-R01}. Hence, the reduction in domain size (i.e., global to boundary domain) did not lead to higher computational efficiency. 

\subsection{Overall performance} 
\label{sec:overall_performance}

\subsubsection{Performance against climatology and synoptic observations}

\begin{figure}[H]
    \centering 
    \includegraphics[trim = 3mm 4mm 4mm 4mm, clip, width=\textwidth]{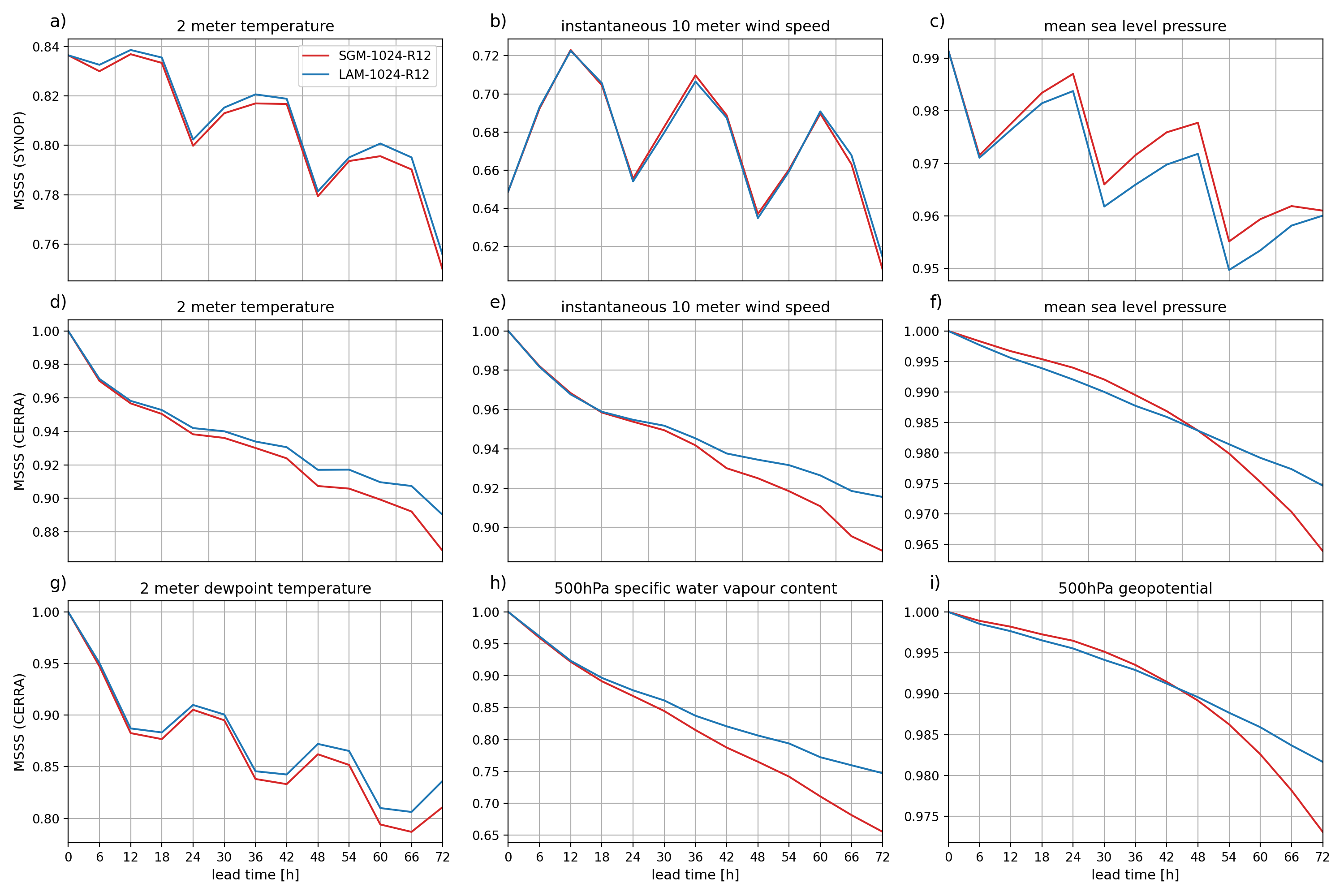}
    \caption{Overall performance of \texttt{LAM-1024} and \texttt{SGM-1024} with rollout training for 2020. (a--c) MSSS compared to the CERRA climatology verified against SYNOP observations. (d--i) MSSS compared to the CERRA climatology verified against CERRA reanalysis for a selection of variables. Higher MSSS values indicate better skill. Reanalyses are used as initial conditions for both SGM and LAM, and also serve as boundary forcings for the LAM (\texttt{ideal} inference setup). Results are displayed for forecast lead times extending up to +72 hours. Forecasts are initialised on 00 UTC each day.} 
\label{fig:fig_R1_MSSS}
\end{figure}

In the following, verification scores for the \texttt{ideal} inference setup, i.e., utilizing reanalyses as initial conditions and boundary forcings (see subsection~\ref{subsubsec:ideal}), are presented. Fig.~\ref{fig:fig_R1_MSSS} shows the Mean Squared Skill Score (MSSS) -- with respect to the CERRA climatology -- verified against SYNOP observations (panels a--c) or CERRA reanalysis (d--i), for \texttt{LAM-1024-R12} and \texttt{SGM-1024-R12} models. These two models were selected as they demonstrated the best performance based on training and validation losses and are thus most suitable for an overall performance assessment. In general, both LAM and SGM produce skilful forecasts with respect to climatology for lead times up to three days, as can be seen for a selection of variables (Fig.~\ref{fig:fig_R1_MSSS}). When verifying against SYNOP observations, LAM and SGM show similar performance at both short and later lead times (Figs.~\ref{fig:fig_R1_MSSS}a--c), although a slight advantage of SGM is observed for mean sea level pressure (msl, Fig.~\ref{fig:fig_R1_MSSS}c). When verifying against CERRA reanalysis (Figs.~\ref{fig:fig_R1_MSSS}d--i), performance remains generally similar at short lead times. However, at later lead times, the LAM outperforms the SGM across all variables, even though the performance of both models declines (Figs.~\ref{fig:fig_R1_MSSS}d--i). This slower decrease in performance for LAM can be attributed to the reanalysis dataset providing ideal boundary forcings, as already mentioned in subsection~\ref{subsubsec:ideal}. This will be discussed in more detail in Section~\ref{sec:discussion}.

\begin{figure}[H]
    \centering
    \includegraphics[width=0.9\linewidth]{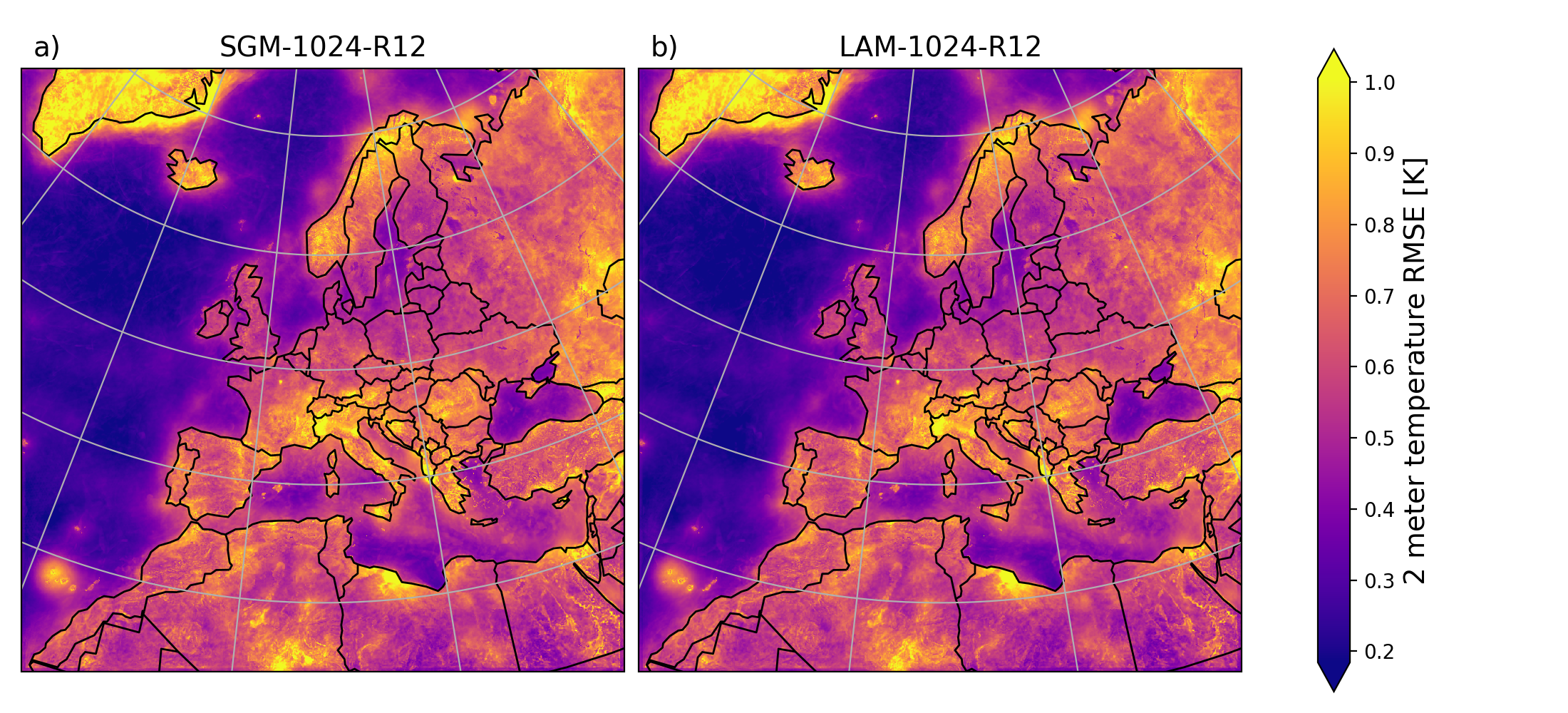}\\
    \includegraphics[width=0.9\linewidth]{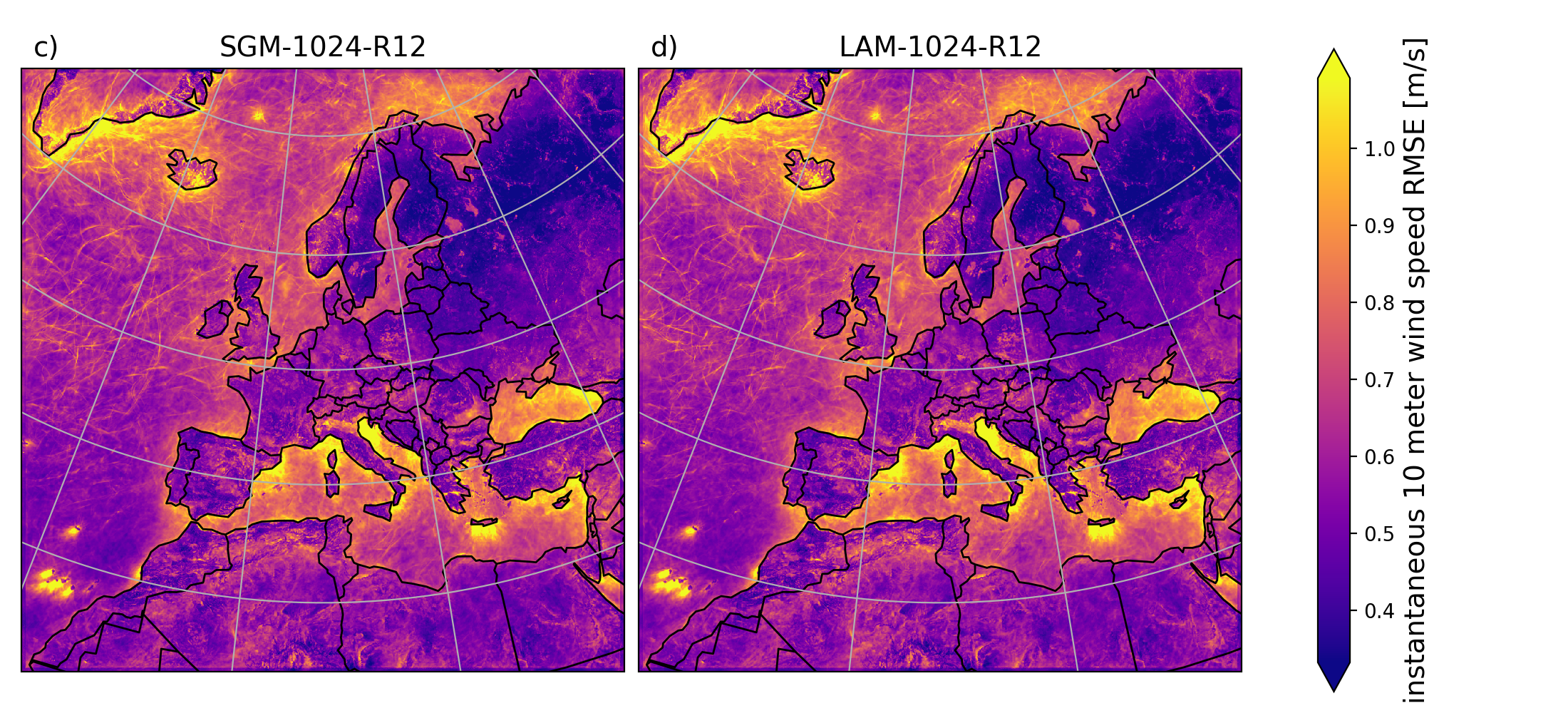}
    \caption{Spatial distribution of RMSE with respect to reanalysis over the regional domain for 2 meter temperature (a, b) and 10 meter wind speed (c, d) with \texttt{SGM-1024-R12-ideal} (a, c) and \texttt{LAM-1024-R12-ideal} (b, d). Results are displayed for the forecast lead time of +6 hours, with forecasts initialised on 00 UTC.}
    \label{fig:spatial-rmse}
\end{figure}

Fig.~\ref{fig:spatial-rmse} shows the spatial distribution of forecast Root Mean Squared Error (RMSE) values throughout the regional domain. Given the close resemblance of the results for SGM and LAM one can conclude that the errors are mainly dominated by inherent difficulty in forecasting rather than the model type used. Discussing the origins of the identifiable spatial error patterns is, however, beyond the scope of this paper.

\subsubsection{Activity and extremes}
\begin{figure}[H] 
    \centering 
     \includegraphics[trim = 4mm 4mm 4mm 4mm, clip, width=\textwidth]{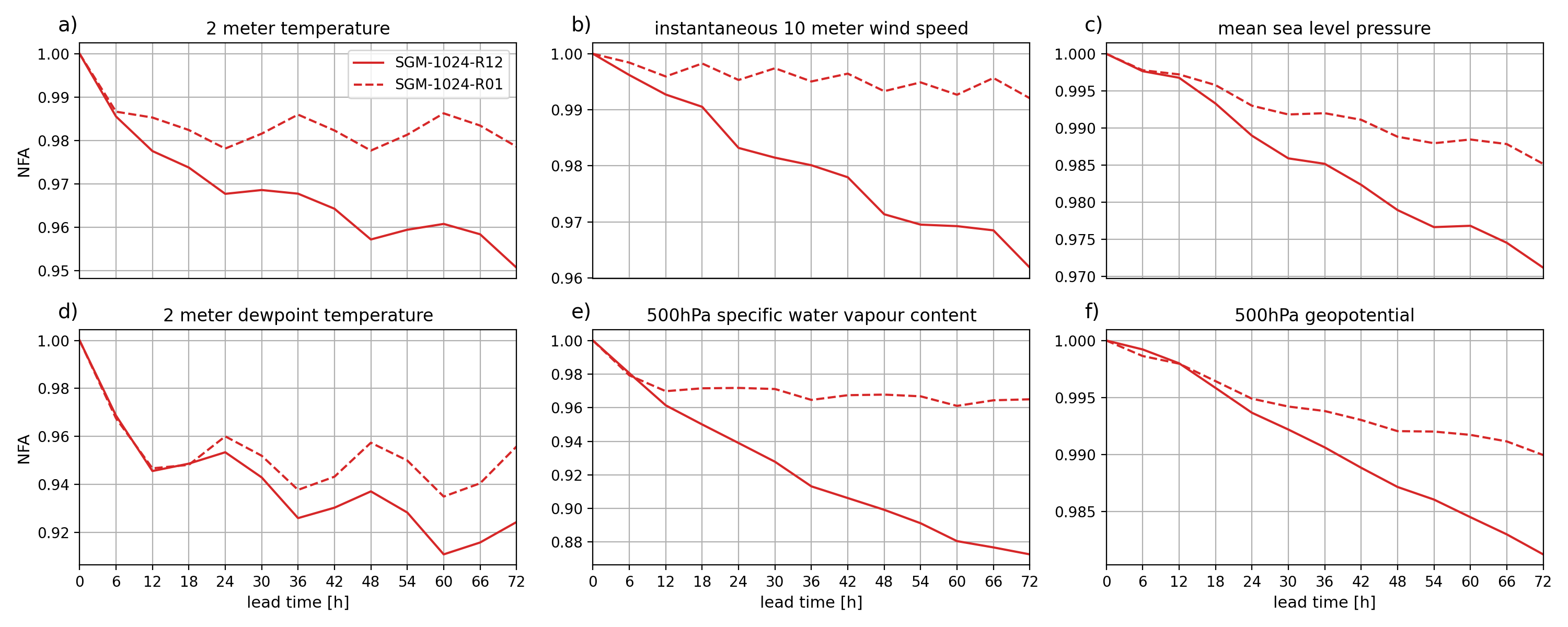}
    \caption{Normalised forecast activity as 
a function of lead time for \texttt{SGM-1024} without (dashed) and with (solid) rollout training for a selection of variables (\texttt{ideal} inference setup). Values closer to 1 indicate that the forecast activity aligns more closely with that of the CERRA reanalysis. All forecasts included here are initialised at 00 UTC from reanalysis.} 
    \label{fig:activity}
\end{figure}

Forecast activity, defined as the standard deviation of the forecast anomaly with respect to CERRA climatology (see also Appendix~\ref{app:metrics}), is assessed to evaluate variability of forecasts. Fig.~\ref{fig:activity} presents the normalised forecast activity (NFA) per lead time, comparing the forecast activity to CERRA reanalyses for \texttt{SGM-1024} forecasts (\texttt{ideal} inference setup), both with and without rollout training. As clearly illustrated, and consistent with prior observations in global MLWP models \cite[e.g.,][]{bouallegue2024}, rollout training leads to a significant decrease in activity for all variables compared to the CERRA reanalysis. This suggests a decrease in the standard deviation of forecast anomalies, resulting in an under-representation of the atmosphere's real spatial variability. It thus implies a loss of realism of the forecasts at later lead times when using rollout training. Similar results were observed for \texttt{LAM-1024}, as well as both \texttt{SGM-512} and \texttt{LAM-512} configurations (not shown). 

\begin{figure}[H] 
    \centering 
    \includegraphics[trim = 3mm 3mm 3mm 3mm, clip, width=0.85\textwidth]{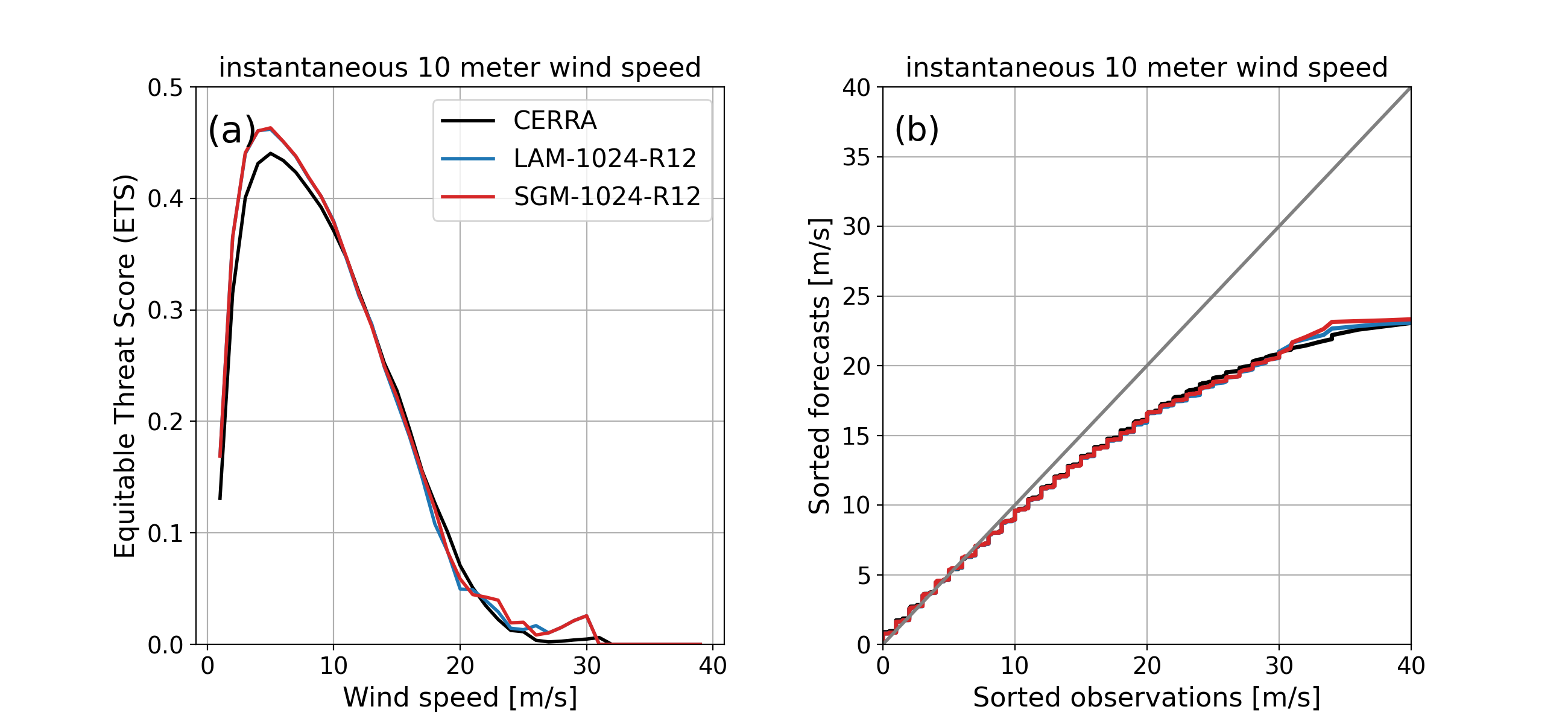}
    \caption{Performance for wind speed extremes. (a) Equitable threat score and (b) quantile-quantile plot against synoptic observations for CERRA, as well as \texttt{LAM-1024} and \texttt{SGM-1024} with rollout training. Results are presented for the forecast lead time of +6 hours. Reanalyses are used as initial conditions for both SGM and LAM, with forecasts initialised at 00 UTC.} 
\label{fig:ETS_plot_lt6}
\end{figure}

Regarding model performance for extremes with respect to SYNOP observations, \texttt{LAM-1024-R12} and \texttt{SGM-1024-R12} exhibit almost identical results at a lead time of 6h (Figs.~\ref{fig:ETS_plot_lt6}a--b). Furthermore, their performance is very similar to that of the CERRA reanalysis, something which is noticeable in both the Equitable Threat Scores (ETS) calculated across various wind speed thresholds (Fig.\ref{fig:ETS_plot_lt6}a) and when examining the complete distribution through a quantile-quantile plot (Fig.\ref{fig:ETS_plot_lt6}b). It should be noted that CERRA contains weaker wind events than those actually observed (Fig.~\ref{fig:ETS_plot_lt6}b), which explains the underestimation also found in LAM and SGM.

\subsection{Inter-model comparison}

\subsubsection{Performance per variable}
\label{sec:performance_per_variable}

\begin{figure}[H]
    \centering
    \includegraphics[trim = 4mm 4mm 15mm 9mm, clip, width=\textwidth]{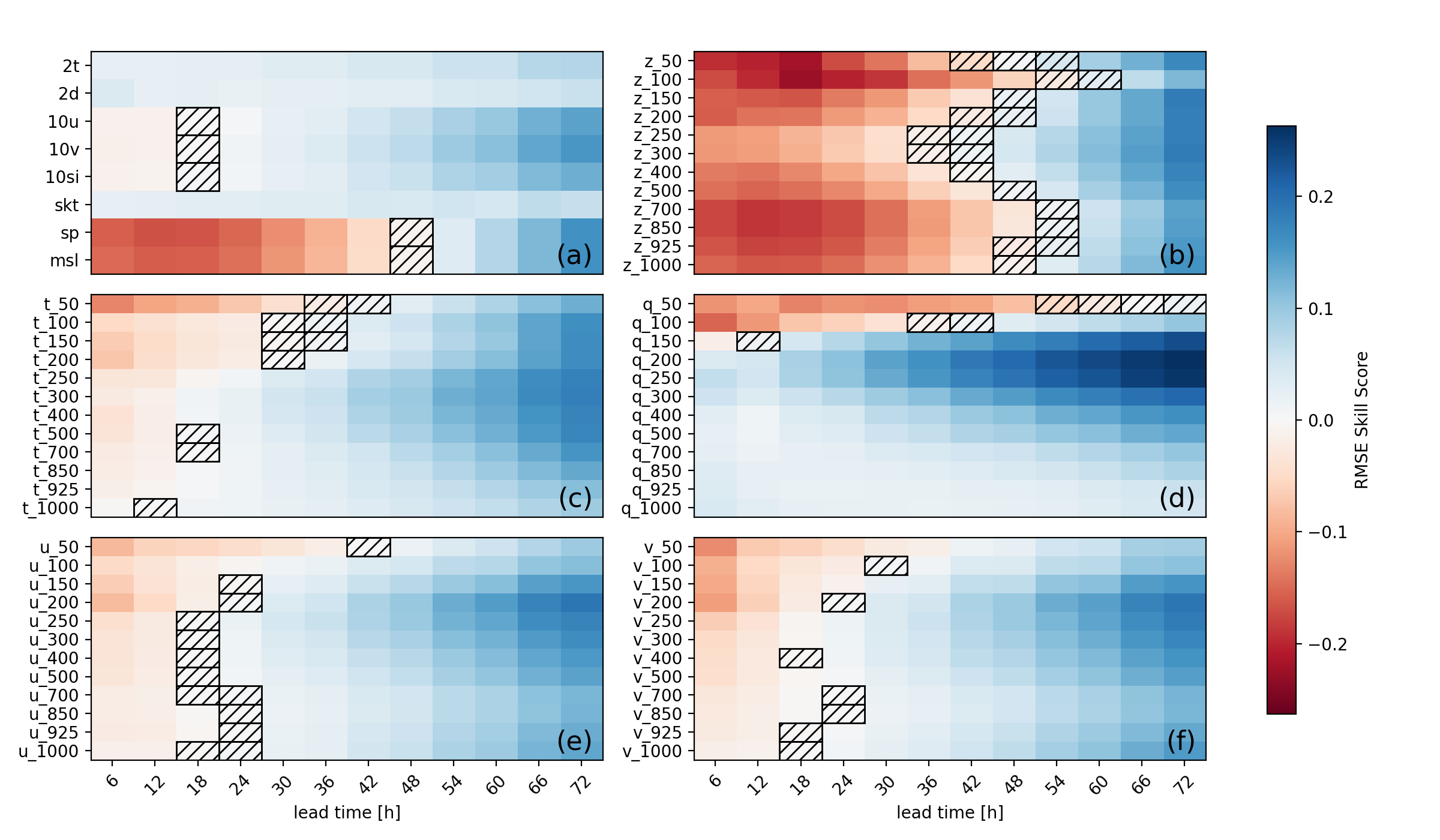}
    \caption{RMSE skill scores for \texttt{LAM-1024-R12} with respect to \texttt{SGM-1024-R12}. The scorecard compares RMSE against the CERRA dataset between LAM and SGM across a selection of variables and pressure levels for lead times ranging from 6 to 72 hours. Blue (resp., red) indicates that LAM is better (resp., worse) than SGM, while hatched cells indicate no significant difference. Reanalyses are used as initial conditions for both SGM and LAM, and also serve as boundary forcings for the LAM (\texttt{ideal} inference setup). Forecasts are initialised at 00 and 12 UTC.}
    \label{fig:fig_RMSE_ScoreCard_1024}
\end{figure}

The performance of LAM and SGM with respect to CERRA varies depending on the specific variables considered. Fig.~\ref{fig:fig_RMSE_ScoreCard_1024} provides a high-level overview of the RMSE skill score relative to CERRA between LAM and SGM for all variables for the \texttt{ideal} inference setup. The SGM generally exhibits better performance at a lead time of 6 hours for a wide range of variables. More specifically, SGM outperforms LAM at the lead time of 6 hours for variables with synoptic-scale variations, hereafter referred to as synoptic-scale variables, such as surface pressure (sp) and msl (Fig.~\ref{fig:fig_RMSE_ScoreCard_1024}a) and geopotential at all pressure levels (z, Fig.~\ref{fig:fig_RMSE_ScoreCard_1024}b). Interestingly, the LAM performs slightly better than the SGM for smaller-scale variables such as 2 meter temperature and specific humidity at high pressure levels (Figs.~\ref{fig:fig_RMSE_ScoreCard_1024}a and d). Note that the initially observed superior relative performance of SGM reduces over time when using reanalyses as boundary forcings, since LAM benefits from this ideal forcing data. Indeed, the LAM significantly outperforms the SGM at later lead times for most variables (e.g., for temperature t, specific humidity q, u and v component of wind, Figs.~\ref{fig:fig_RMSE_ScoreCard_1024}c--f). Nevertheless, the SGM tends to demonstrate better relative performance in the upper atmosphere, as illustrated for temperature and specific humidity in Figs.~\ref{fig:fig_RMSE_ScoreCard_1024}c and d.

\subsubsection{Model size}
\label{sec:model_scaling}

\begin{figure}[H] 
    \centering
    \includegraphics[trim = 4mm 4mm 15mm 9mm, clip, width=\textwidth]{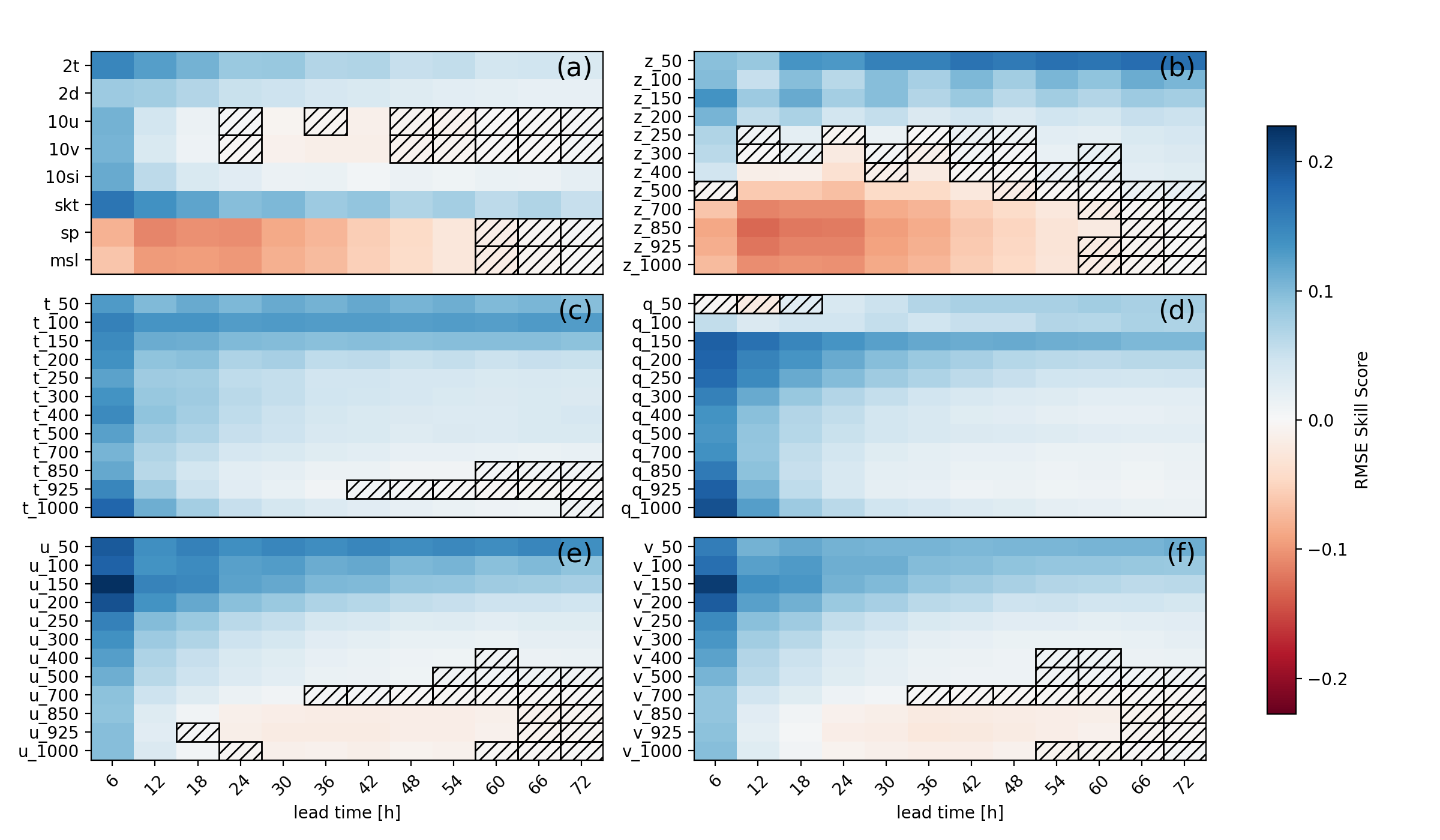}
    \caption{Influence of the number of model parameters on performance. Same as Fig.~\ref{fig:fig_RMSE_ScoreCard_1024} but comparing \texttt{SGM-1024-R01} to \texttt{SGM-512-R01}.} 
    \label{fig:fig_RMSE_ScoreCard_1024vs512}
\end{figure}

Model size, determined here by the number of channels, has a substantial impact on forecast performance. For example, Fig.~\ref{fig:fig_RMSE_ScoreCard_1024vs512} shows RMSE skill scores for \texttt{SGM-1024-R01} compared to \texttt{SGM-512-R01} for the \texttt{ideal} inference setup. This comparison entails an increase in parameter count from 62 million to 246 million trainable parameters. Generally, there is a significant improvement in terms of RMSE for both surface and pressure level variables. An exception are the synoptic-scale variables closer to the surface (e.g., msl and sp,  Fig.~\ref{fig:fig_RMSE_ScoreCard_1024vs512}a and geopotential, Fig.~\ref{fig:fig_RMSE_ScoreCard_1024vs512}b). Similar results are obtained for LAM and models with rollout (not shown).

\subsubsection{Ideal vs. operational settings}
\label{sec:ideal_operational}

\begin{figure}[H] 
    \centering
    \includegraphics[trim = 4mm 4mm 4mm 4mm, clip, width=\textwidth]{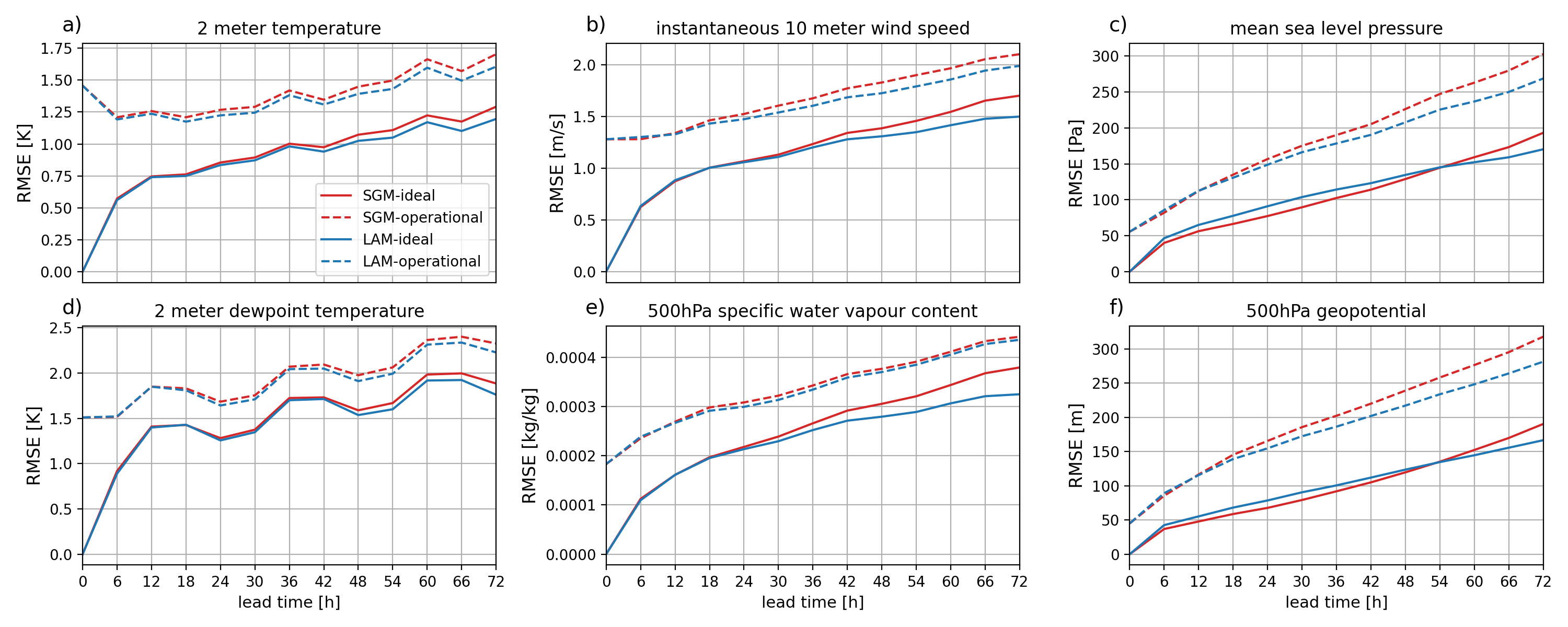}
    \caption{Influence of initialisation and boundary forcing datasets on performance, showing RMSE with respect to the CERRA reanalysis as a function of lead time for selected variables. Scores are presented for \texttt{LAM-1024-R01} and \texttt{SGM-1024-R01} using different initialisation datasets: reanalyses from ERA5 and CERRA (\texttt{ideal}, solid) and operational data from IFS-HRES analyses and forecasts (\texttt{operational}, dashed). Forecasts are initialised at 00 UTC.} 
    \label{fig:fig_R5_od_ea}
\end{figure}

Here, verification scores under the \texttt{ideal} inference setting (initialisation and boundary forcing from reanalysis, see Section~\ref{subsubsec:ideal}) are compared with those of the \texttt{operational} inference setup that more closely resembles operational settings (initialisation from analysis, boundary forcing by an external forecast, see Section~\ref{subsubsec:operational}). Inference results using operational data exhibit higher RMSE compared to those based on reanalysis datasets across various variables for both LAM and SGM (Fig.~\ref{fig:fig_R5_od_ea}). This is mainly due to  an initial error at a lead time of 0 hours, reflecting how much the interpolated IFS analysis deviates from the CERRA reanalysis, which was selected as the reference for verification. Variations in error growth over lead times are observed between the LAM and SGM in operational-like settings (e.g., for geopotential, Fig.~\ref{fig:fig_R5_od_ea}f), despite their close similarity at shorter lead times. In both inference settings, the LAM demonstrates better performance compared to the SGM at later lead times.

\subsubsection{Influence of the boundary forcings on LAM performance}
\label{sec:boundary_forcings}

\begin{figure}[H] 
    \centering
    \includegraphics[trim = 4mm 4mm 4mm 4mm, clip, width=\textwidth]{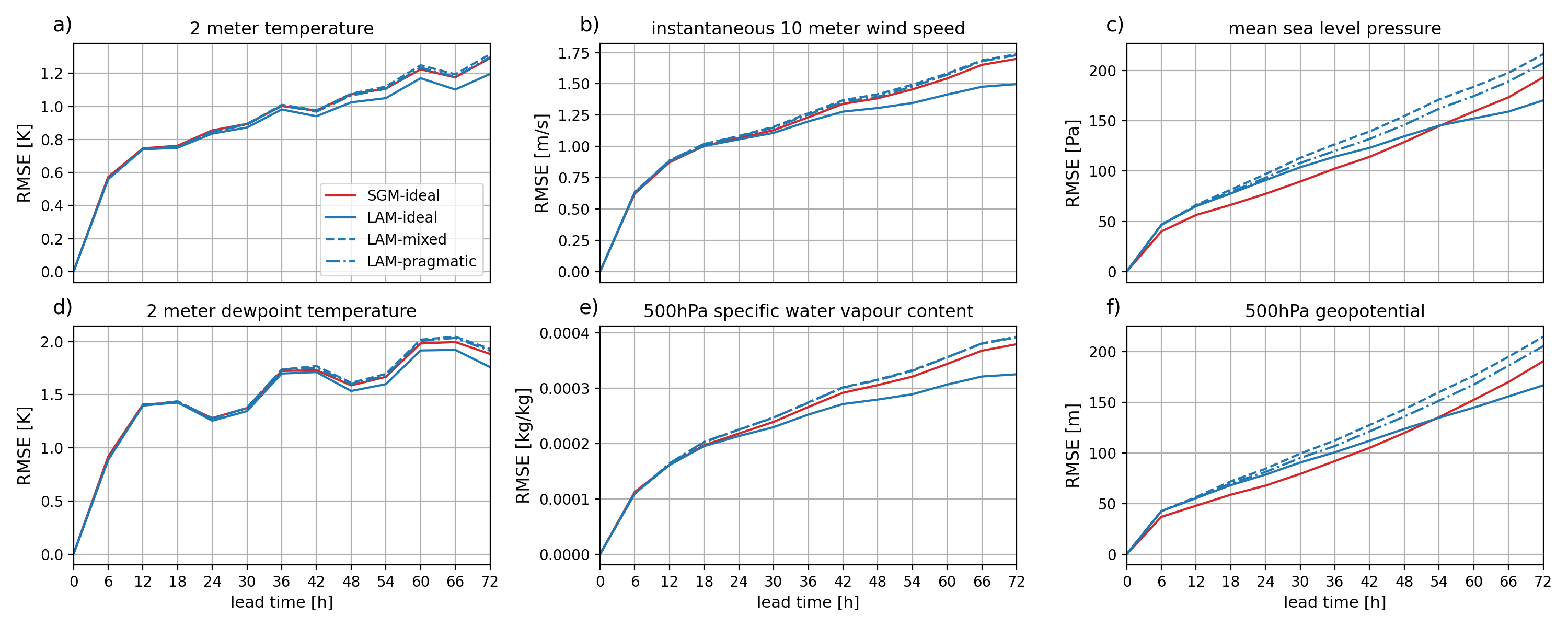}
    \caption{Same as Fig.~\ref{fig:fig_R5_od_ea} but for \texttt{LAM-1024-R01} using different forcing datasets: ERA5 reanalyses (\texttt{ideal}, solid), operational IFS-HRES reduced to O96 resolution (\texttt{pragmatic}, dash-dot) as well as SGM forecasts (\texttt{mixed}, dashed) on the LAM boundary domain. For comparison, results for \texttt{SGM-1024-R01} initialised with reanalyses dataset are also presented.} 
    \label{fig:LAM_forcing}
\end{figure}

Inherent to the LAM approach is the flexibility to force it with a preferred set of boundary forecasts. Here, after initialisation from reanalysis, we compare three types of boundary forcings during rollout inference: \texttt{ideal} (i.e., using ERA5 reanalysis), \texttt{pragmatic} (i.e., using IFS-HRES NWP forecasts) and \texttt{mixed} (i.e., using the global part of the SGM). Fig.~\ref{fig:LAM_forcing} shows that, especially at later lead times, the  \texttt{ideal} forcings result in superior performance for all variables, as expected. The fact that the \texttt{pragmatic} forcings slightly outperform the \texttt{mixed} ones indicates that, at least for the LAM, the boundary forecasts of SGM provide it with less valuable information than the IFS forecasts. This contrasts with the observation that the SGM forecasts with the same initialisation on the regional domain, shown in the same figure for reference, outperform the LAM forecasts with both \texttt{mixed} and \texttt{pragmatic} forcings (e.g., for msl and geopotential, Figs.~\ref{fig:LAM_forcing}c and f). This suggests that the SGM is more proficient than the LAM in utilizing its own global domain forecasts to make predictions in the regional domain. 

\begin{figure}[H] 
    \centering
    \includegraphics[trim = 4mm 4mm 4mm 4mm, clip, width=\textwidth]{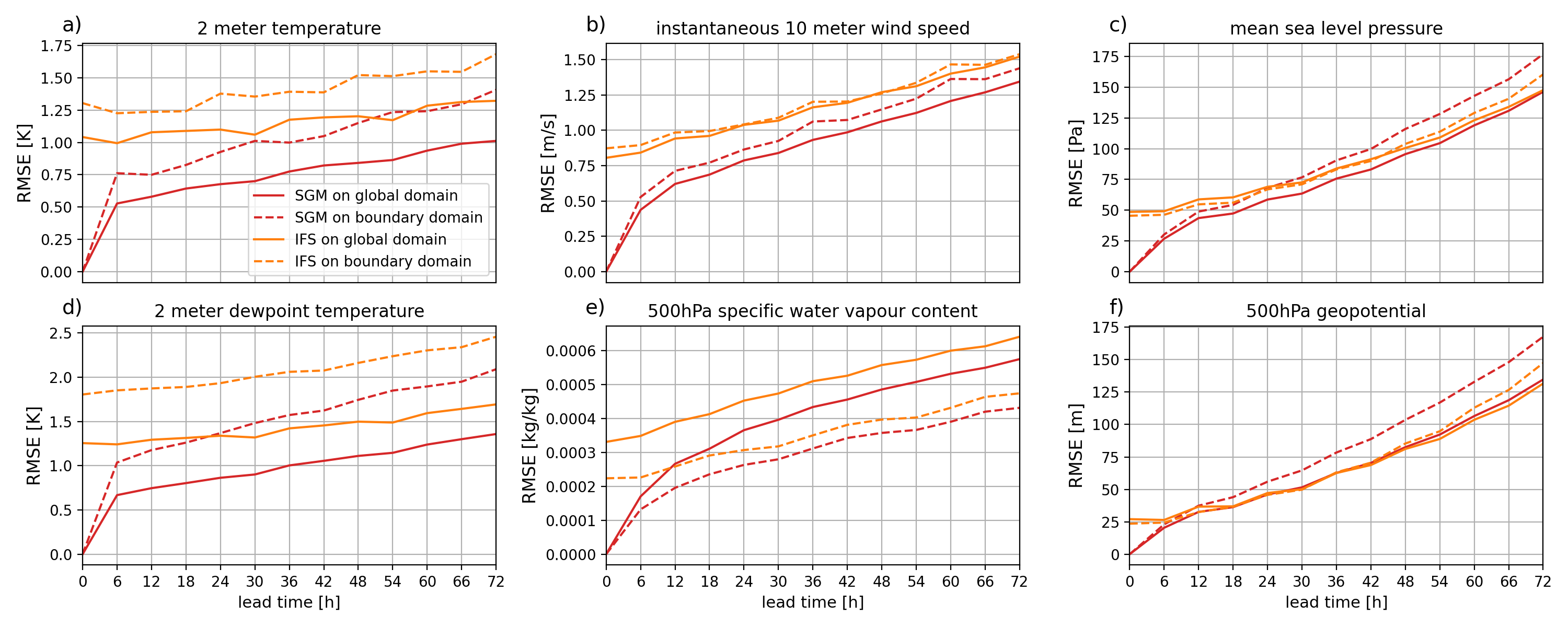}
    \caption{RMSE scores against ERA5 over the global domain (i.e. globe without regional domain, solid) and the boundary domain (dashed) for \texttt{SGM-1024-R01-ideal} (red) and operational IFS reduced to O96 resolution (orange). Forecasts are initialised at 00 UTC.} 
    \label{fig:boundary_plot}
\end{figure}

To investigate the relationship between boundary quality and LAM results, Fig.~\ref{fig:boundary_plot} shows the performance of the IFS and SGM on the global and boundary domains against ERA5. IFS tends to show a slower error growth than the SGM, for example for 2 meter temperature and 10 meter wind speed (Figs.~\ref{fig:boundary_plot}a and b). This result is not surprising since this is a SGM without rollout training. More importantly, for msl and 500hPa geopotential the difference between IFS and SGM is larger on the boundary domain than on the global domain (Figs.~\ref{fig:boundary_plot}c and f). This explains the relative performance difference between the LAM forced by these two boundary forecasts (\texttt{pragmatic} and \texttt{mixed} setups) as shown in  Figs.~\ref{fig:LAM_forcing}c and f. Further interpretation of these results can be found in Section~\ref{sec:discussion}.

\subsubsection{Generalisability to unseen times of day}
\begin{figure}[H] 
    \centering
    \includegraphics[width=0.49\linewidth]{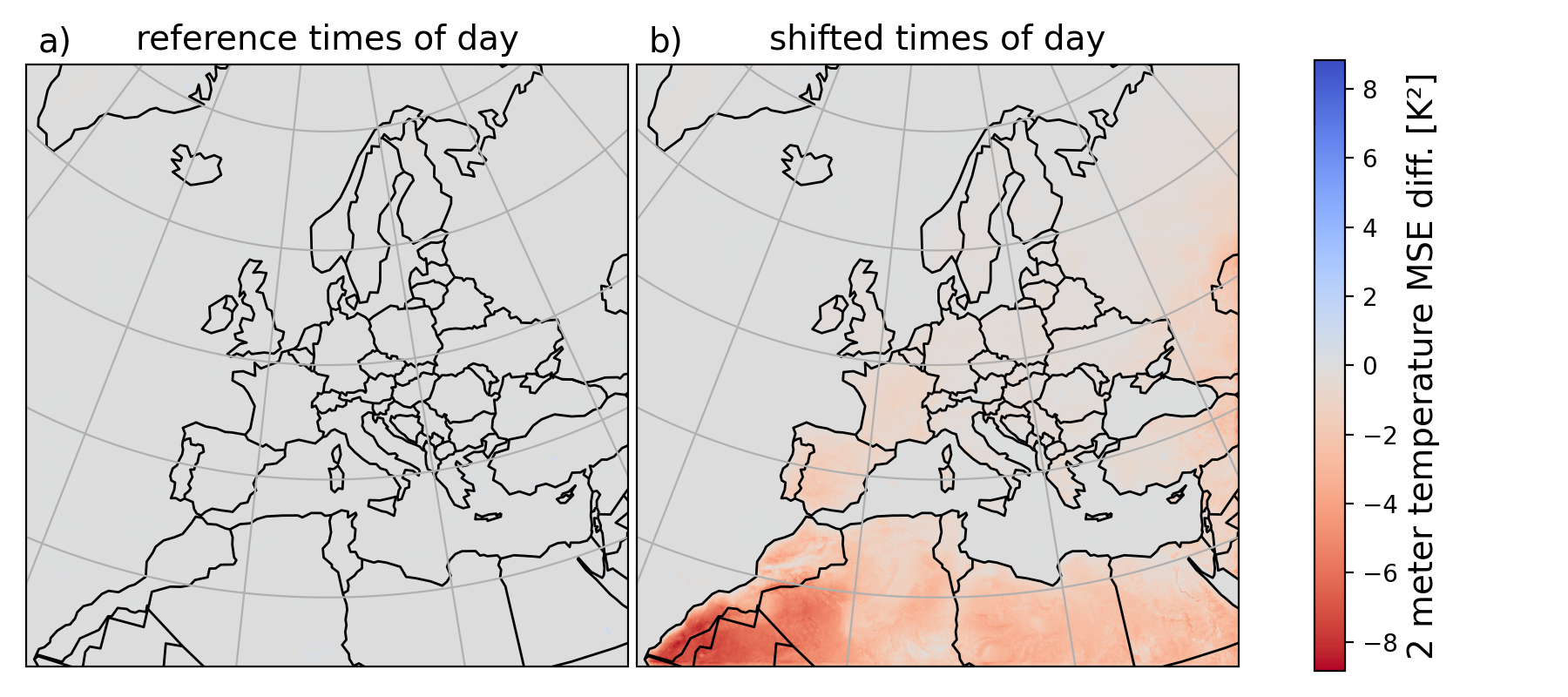}
    \includegraphics[width=0.49\linewidth]{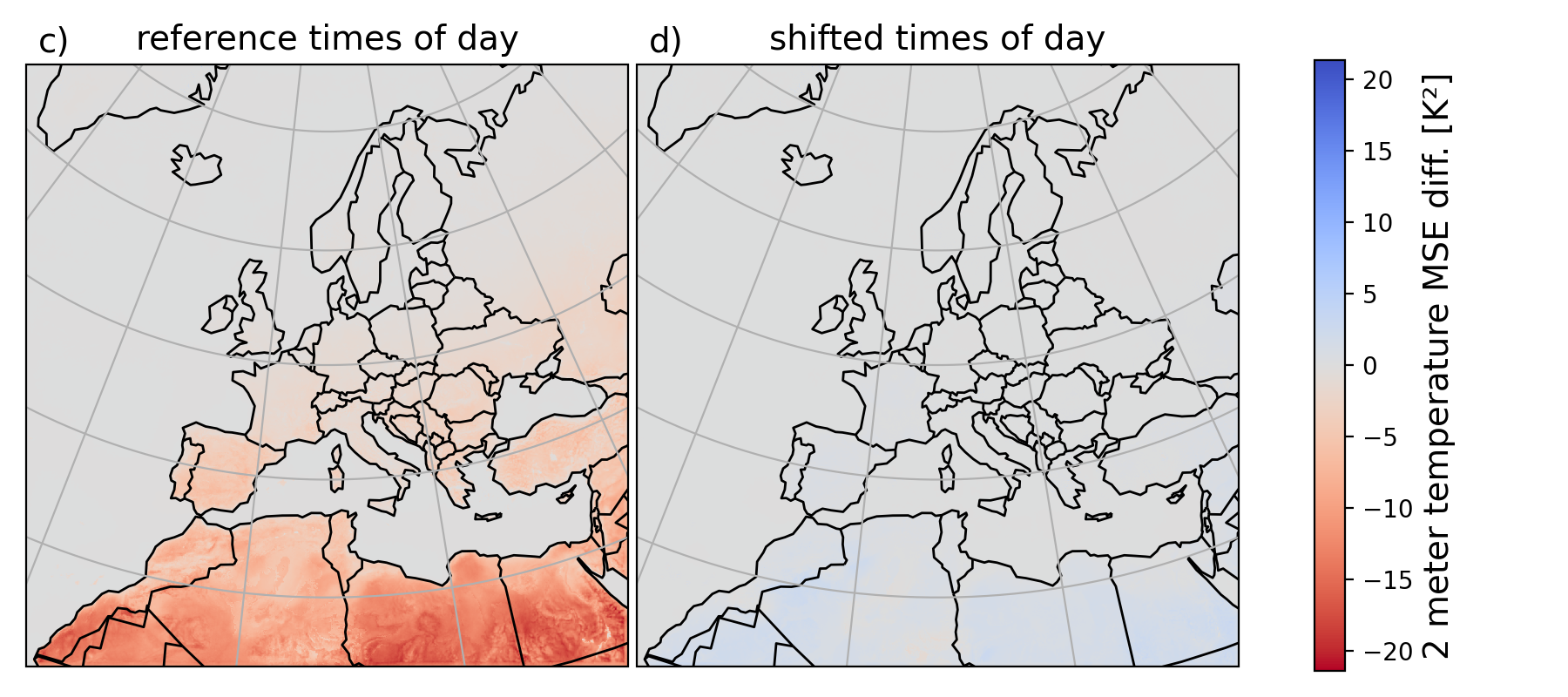}
    \caption{Difference in MSE for 2 meter temperature between \texttt{SGM-512} and \texttt{LAM-512} over the period 2020. Red (resp., blue) indicates better performances for SGM (resp., LAM). In panels (a, b) the comparison is made between models trained on the reference dataset and initialised at (a) reference and (b) shifted times of day. For panels (c, d) only the LAM is retrained on the shifted dataset, and both models are initialised at (c) reference and (d) shifted times of day.} 
    \label{fig:generalisability}
\end{figure}

In this experiment, described in Section~\ref{sec:generalisability}, forecasts were initialised both on reference (00, 06, 12 and 18 UTC) and shifted (03, 09, 15 and 21 UTC) times of day. In the standard training procedure (see Section~\ref{subsec:training}), the reference times of day are seen by the models during training, while the shifted times remain unseen. A striking difference in the relative performance of SGM and LAM between seen and unseen times of day is found for temperature-related variables close to the surface. This difference is illustrated in Figs.~\ref{fig:generalisability}a and b for 2 meter temperature forecasts.

At unseen times of day, the SGM outperforms the LAM significantly, with the differences localised mainly over Northern Africa (Fig.~\ref{fig:generalisability}b). In contrast, at times of day present in the training data, both models perform essentially the same over the whole regional domain (Fig.~\ref{fig:generalisability}a). Results are similar when changing to 1024-channel models (not shown). Further, an alternative LAM was trained on the shifted times of day only. The performance of SGM compared to this alternative LAM on 2 meter temperature forecasts is shown in Figs.~\ref{fig:generalisability}c--d. It can be observed that the LAM specifically trained on shifted times of day slightly outperforms the SGM for these times, as the SGM was not retrained (Fig.~\ref{fig:generalisability}d). However, it now highly underperforms compared to SGM for the reference times of day (Fig.~\ref{fig:generalisability}c). The origin of this difference in temporal generalisability will be further discussed in Section~\ref{sec:insights}.

\section{Discussion}
\label{sec:discussion}

\subsection{Performance differences}
\label{subsec:performance_differences}

In this study, various differences have been observed between LAM and SGM results. For example, the scorecard in subsection~\ref{sec:performance_per_variable} indicates that the model performance of \texttt{SGM-1024-R12-ideal} is better than \texttt{LAM-1024-R12-ideal} at higher altitudes, for both small-scale and synoptic-scale variables. An explanation could be that variables with small-scale variations closer to the surface tend to smoothen towards more large-scale features at higher altitudes, due to the lack of topographic influences, and \texttt{SGM-1024} is particularly effective with such variables (as illustrated by z and msl, Fig.~\ref{fig:fig_RMSE_ScoreCard_1024}). That effectiveness might originate from the additional exposure to such large-scale features on SGM's global domain and which it is incentivised during training to forecast well. The model has the capacity to do so through the long edges in the global domain that capture long-distance information. In addition, part of LAM's relative underperformance for these type of variables and regimes could be due to the rapid movement of weather systems at high altitudes driven by increased wind speeds. Considering a larger boundary domain for LAM might improve its ability to capture the dynamics of faster-moving weather patterns. 

At later lead times, better performance was observed for LAM in both the ideal and operational-like inference setups (Fig.~\ref{fig:fig_R5_od_ea}). It is easy to understand this in the ideal setting, where the LAM is forced by reanalysis rather than forecasts (which is not feasible during operational forecasting). The effect is quite more subtle and nuanced in the operational-like setting. In Fig.~\ref{fig:LAM_forcing}, it can be observed that LAM forced with IFS forecasts outperforms a forcing with the SGM global forecasts. This indicates that on the boundary region of the LAM, the quality of the IFS forecasts is somewhat higher than that of the SGM. However, Fig.~\ref{fig:boundary_plot} reveals that this is only the case for a few variables, such as mean sea level pressure and 500 hPa geopotential. In addition, for these variables the relative difference between SGM and IFS is much larger on the boundary domain of LAM than on the full global domain. This suggests that accurate forecasting on the boundary region is challenging for the SGM, which is not unexpected given its transitional nature. Still, when comparing forecasts initialised from reanalysis (Fig.~\ref{fig:fig_R5_od_ea}), SGM outperforms LAM forced by SGM global forecasts, as well as LAM forced by IFS. This shows that any deficiencies in the SGM forecasts outside the regional domain, if indeed they are present and significant, can be compensated for by the SGM to forecast with high accuracy on the regional domain.

The observation in Fig.~\ref{fig:LAM_forcing} that \texttt{SGM-1024-R01} outperforms LAM with IFS forcing when both are initialised from reanalysis but not when both are initialised from (partially interpolated) operational analysis, suggests a reason for the underperformance of SGM at later lead times in the operational-like setting. Specifically, it could be due to an initial error, induced by initial conditions not encountered during training, which grows throughout the SGM forecast. A similar effect would be less severe for LAM, since it has IFS forcings which are free of such error and thus moderate the growth of any initial errors. In view of this, an important caveat needs to be made. In the operational-like inference setup, models were used without rollout training and without additional fine-tuning to adapt them to the new initialisations and forcings used. Both types of additional training, which can be combined in a single phase \cite{IFSv1}, would likely reduce the initialisation error and suppress its growth over time. This could potentially change the relative performance at later lead times presented in subsections~\ref{sec:ideal_operational} and~\ref{sec:boundary_forcings}.

\subsection{Challenges and limitations}
Since the goal of this study is the comparison of LAM and SGM approaches, optimising either of them individually to achieve the best possible performance was determined to be out of scope. For example, in the operational experiments (see Section~\ref{subsubsec:operational}), the models have not been fine-tuned on operational data from forecasting systems. Therefore, the model has no knowledge of the differences between reanalysis and operational data. This final fine-tuning step of the models, together with a rollout phase, would be important to ensure the usability of either model for actual operational forecasting. It is also the reason why in this work the MLWP models have not been directly compared to operational NWP models. Optimisations particular to only one of the two methods, such as the use of future boundary forcings for LAM \cite{adamov2025}, have not been considered as they would skew a comparison. 

In this study, LAM and SGM are trained to emulate the CERRA reanalysis within the regional domain. The accuracy of their predictions is thus necessarily linked to the quality of the CERRA dataset, encompassing its strengths, but also its limitations. One of the limitations of CERRA has been observed in Fig.~\ref{fig:ETS_plot_lt6}, which reveals a strong underestimation of wind speed extremes with respect to stations observations. This underestimation is a consequence of the inherent averaging effects of the CERRA grid system. Consequently, both LAM and SGM exhibit an underestimation of local extremes consistent with what is obtained for CERRA. While it is well-established that deterministic MSE-based MLWP systems tend to underestimate extremes \cite[e.g.,][]{nipen2024, adamov2025}, such a conclusion cannot be inferred within the context of our study. Although this might be related to the comparatively lower spatial resolution of CERRA, evaluating the relative performances of LAM and SGM for extremes using a higher-resolution reanalysis is beyond the scope of this study. Similarly, investigating the suitability and relative performance of LAM and SGM in terms of physical consistency, capabilities on small training datasets, and probabilistic forecasting \cite[e.g., ][]{larsson2025diffusionlam} is left for future work. The latter could, in addition to providing uncertainty estimates, enhance forecast activity.  

\subsection{Insights}
\label{sec:insights}
SGM trained slightly faster (~10\% improvement) than LAM, indicating a higher computational efficiency. Since this depends on the resolution of the global and boundary domains, it will change as this resolution is increased. Still, it does show that the traditional computational advantages of LAM in an NWP context do not translate directly to a ML context. Further, although LAM and SGM have the same number of parameters, SGM uses this capacity to forecast on both a low-resolution global scale as well as on the high-resolution regional domain. The LAM can outsource this computational complexity to an external global model and use the full parameter capacity to specialise for the regional domain. This means that, when accounting for the complexity required to provide boundary forcings, the total computational complexity of LAM is higher than that of the all-in-one SGM approach. However, it also implies that LAM can benefit from higher quality global models for boundary forcing, which especially impacts predictions at later lead times. The interplay of initialisation, boundary forcing and global forecasting is subtle however, as already discussed in detail in Section~\ref{subsec:performance_differences}.

Further, our study has found that the configuration of LAM and SGM architectures can have a major impact on model performance. For example, when designing the LAM architecture, one of the key features is the configuration of the boundary domain. This includes choosing its extent, the resolution of data and hidden grids over this domain and how they are connected through edges in the encoder. After exploring and testing various configurations, a key insight for designing LAM graphs was that while the size of the boundary does impact performance, it is even more important to ensure that sufficient hidden nodes are connected exclusively to boundary nodes. It was achieved here by having the hidden mesh extend deep into the boundary zone at high resolution (see Section~\ref{subsec:models}). However, this also caused a decrease in training efficiency compared to the SGM. Further improvements to the LAM graph may be possible. For example, lowering the resolution of the hidden grid over the boundary domain would improve computational efficiency, while hopefully preserving its ability to absorb boundary information.

For SGM there is the possibility to adjust the loss contributions of the regional and global domains using a hyperparameter. Initially, these components were weighted based on the geographical size of the domain. This resulted in a 6.7\% weighting for the regional domain. However, experimentation showed that an increased weighting led to better results inside the regional domain for shorter lead times (i.e., 25\% regional domain weighting was used in all experiments). Note that with a regional loss weighting of 100\% the SGM converges towards a LAM with poor boundary conditions. Therefore, a balanced solution is required; we are interested in good regional performance, yet a loss of performance in the global domain eventually leads to reduced regional performance at later lead times. Hence, lowering the weight on the regional domain may lead to improved performance of the SGM at later lead times and impact the comparison reported in subsections~\ref{sec:ideal_operational} and \ref{sec:boundary_forcings}.

In Section~\ref{sec:generalisability}, it was observed that the SGM outperforms LAM in generalizing the forecasting of near-surface temperatures over Northern Africa to unseen times of day (Fig.~\ref{fig:generalisability}). This could indicate that the LAM is missing part of the physics related to heating and cooling of some arid regions at specific times of day. On the contrary, the SGM seems to have learned part of these physical relationships, despite not being trained on data from these regions at these specific times of day. The SGM could potentially learn this from comparable geographical areas in other locations around the globe. Since the SGM predicts both regional and global domains, its training data includes the full 24-h daily cycle, continuously laid out over the different geographical time zones, although mostly at a low resolution. The LAM, however, only has access to training data from the regional and boundary domains. Consequently, it only encounters data for certain types of geography at limited times of day. In terms of temporal generalisation, SGM may therefore benefit from additional training data across multiple time zones. It would also indicate the SGM is capable of generalizing from the low-resolution part of the domain into the high-resolution region.

\subsection{Outlook}
In this study, an overall performance improvement is observed when increasing model size from 62 million parameters (i.e., 512 channels) to 246 million parameters (1024 channels). The smaller models still provide reasonable performance indicating they are suited for explorations in initial model development, before scaling up. The 512-channel models performed especially well on less complex tasks such as modelling of synoptic-scale variables. Consistent improvements have been observed for small-scale variables when scaling up to 1024-channel models, for both LAM and SGM (see subsection~\ref{sec:model_scaling}). However, as often encountered \cite[e.g., ][]{MLPP_informal}, this seems to be at a cost of reduced performance on the synoptic scale. SGM experiments with even more model parameters indicate this further boosts model performance (results not shown). Similarly, developments in natural language processing (e.g., GPT-2/3/4) suggest that more data, parameters and compute could lead to substantial performance increases \cite{openai2024gpt4technicalreport}. When scaling up, it could beneficial for SGM that it has access to more training data compared to LAM due to its training on the full global domain, an aspect also noted in relation to generalisability (see Section~\ref{sec:insights}). This benefit could become more prominent when increasing the resolution of the global domain, as this will make the data additionally available to SGM even larger. For example, increasing the global resolution from O96 to N320 would increase the additional data available to SGM from 3\% to 44\%. Our experiments have also shown that overfitting could become an issue when further increasing parameter count. Therefore, increased regularisation could be a promising direction for future research.

Another difference between LAM and SGM is that SGM has already been used successfully for transfer learning, leveraging the common modelling of the global domain in the transfer learning task. Specifically, by adopting the same resolution and graph structure in the global domain, a large part of the model will be similar and can be reused. This aligns with the practice of building base models with as much high-quality training data as possible, as described above. As the CERRA domain encompasses the forecast regions of many European national weather services, our best SGM experiments (1024 channels with rollout, see Section~\ref{sec:overall_performance}) could be used to initialise a GNN on regional (re)-analyses at higher spatial resolution (e.g., 1--2 km). These regional datasets typically only contain several years of data. Therefore, the 36+ years of CERRA data at 5.5 km resolution could be a promising starting point for a ML model to learn atmospheric dynamics and limit overfitting on a relatively small regional (re)-analysis dataset. Although outside the scope of this study, initial transfer learning experiments have been performed and show promising results. Similar experiments with LAM are planned.

LAM and SGM can potentially be used for other applications than weather forecasting, such as regional climate modelling. The most prominent example of this is the CORDEX Initiative, in particularly the ongoing EURO-CORDEX project \cite{EURO-CORDEX}. Climate ensembles consist of a few dozen different climate simulations of long climate runs of the order of 100 years each with a different earth-system model with its own climate-change signal. Regional simulations using a SGM would require to train and validate outside the regional domain of interest. This would necessitate a large number of GCM model runs in order to properly capture the spread in the climate signal. Another common approach to derive simulations at convection-permitting scales is double nesting with regional models. Since there are no global data available, LAM would be a straightforward choice to generate such climate simulations.

\section{Conclusion}
\label{sec:conclusion}

This study presented the first comprehensive assessment of the relative performance of LAM and SGM for deterministic regional weather forecasting. Using the Anemoi framework, both model types were constructed based on a common architecture with model-specific adaptations and trained in a near-identical setup. Several inference experiments have been performed to investigate differences in terms of model performance under various initialisation and forcing scenarios. By verifying and comparing results against reanalysis, climatology and observations, our study shows that LAM and SGM are competitive ML modelling approaches with similar forecasting performance over the regional domain. Overall forecasting performance is generally good, although both models exhibit an increased smoothing at longer forecast lead times. We have found that SGM slightly outperforms LAM for synoptic-scale variables and shows much greater (temporal) generalisability due to its access to global meteorological data. By focusing solely on the modelling of the regional domain, LAM is able to perform marginally better than SGM for some of the small-scale variables. The different inference experiments show that both SGM and LAM adapt well to a change of initialisation using analysis and forecast data sources, even without additional fine-tuning. In addition, LAM is more flexible in the use of various types of boundary forcings, which can be leveraged to improve forecast performance at later lead times without extensive retraining. To help practitioners choose a model for their data-driven forecasting activities, we have listed the following recommendations. SGM has the advantage of being a more integrated forecasting system than LAM, by being fully self-contained and forecasting on both global and regional domains, which simplifies operationalisation and technical maintenance. SGM also has access to more training data which aids generalisability and can thus be expected to also be valuable when further scaling up model complexity. On the other hand, LAM, with its flexibility in choosing boundary forcings, can swiftly take advantage of improvements in both global NWP and MLWP modelling. LAM can utilize improved global forecasting techniques without the need of extensive retraining and hence saving computational resources, in addition to a marginally better performance on some small-scale variables. It also has its application in contexts where global data might be absent or difficult to acquire (e.g. climate modelling). With careful consideration for the specific implementation challenges inherent to each method, both LAM and SGM could form the basis of a successful deterministic, data-driven weather forecasting system.

\section*{Acknowledgements}
The authors would like to acknowledge ECMWF and the entire Anemoi team for co-development of the Anemoi framework for data-driven weather forecasting. This work was partially supported by the ECMWF Machine Learning Pilot Project (MLPP), a European collaborative initiative to advance data-driven weather prediction, as well as the DE\_330 project of the Destination Earth Initiative, financed by the European Commission. We thank the other participants in these projects for various interesting discussions and feedback during meetings where some of the results included in this work were presented in an earlier form. We thank ECMWF for providing access to the Leonardo EuroHPC supercomputer and to its own HPC infrastructure. We acknowledge LUMI-BE for awarding this project access to the LUMI supercomputer, owned by the EuroHPC Joint Undertaking, hosted by CSC (Finland) and the LUMI consortium through a LUMI-BE Regular Access call. LUMI-BE is a joint effort from BELSPO, SPW Économie, Emploi, Recherche Wallonia, Department of Economy, Science \& Innovation Flanders and Innoviris.

\printbibliography
\newpage
\begin{appendices}
\section{Used variables}
\label{app:used_variables}

\definecolor{MineShaft}{rgb}{0.2,0.2,0.2}
\renewcommand\thetable{A.\arabic{table}} 

\begin{table}[H]
\centering
\caption{Overview of the used variables. Prognostic variables are used as input and output and forcing variables only as input. The derived column indicates whether the variable is absent in the regional (CERRA) or global (ERA5) dataset and needs to be derived from other variables present in the respective dataset.}
\label{tab:variables}
\begin{tblr}{
  cell{2}{2} = {r=5}{fg=MineShaft},
  vline{2-6} = {2-6,8-15}{0.05em},
  vline{3-6} = {17-27}{0.05em},
  hline{1-2,7-8,16-17,27} = {-}{0.08em},
}
Upper air variables              & Pressure levels (hPa)                                      & Derived & Units        & Type       & Normalisation \\
Temperature (t)                  & {50,~100,~150, \\ 200,~250,~300,\\400,~500,~700, \\850,~925,~1000} & No      & K            & Prognostic & mean - std    \\
U component of wind (u)                   &                                                            & No      & m s$^{-1}$   & Prognostic & mean - std    \\
V component of wind (v)              &                                                            & No      & m s$^{-1}$   & Prognostic & mean - std    \\
Specific humidity (q)            &                                                            & CERRA   & kg kg$^{-1}$ & Prognostic & mean - std    \\
Geopotential Height (z)          &                                                            & No      & m            & Prognostic & mean - std    \\
Surface variables                & Height level (m)                          & Derived & Units        & Type       & Normalisation              \\
Skin temperature (skt)           & 0                                                          & No      & K            & Prognostic & mean - std    \\
2m Temperature (2t)              & 2                                                          & No      & K            & Prognostic & mean - std    \\
2m Dewpoint temperature (2d)     & 2                                                          & CERRA   & K            & Prognostic & mean - std    \\
10m u component of wind (10u)             & 10                                                         & CERRA   & m s$^{-1}$   & Prognostic & mean - std    \\
10m v component of wind (10v)        & 10                                                         & CERRA   & m s$^{-1}$   & Prognostic & mean - std    \\
Surface pressure (sp)            & 0                                                          & No      & m s$^{-1}$   & Prognostic & mean - std    \\
Mean sea level pressure (msl)    & Mean sea level                                              & No      & m s$^{-1}$   & Prognostic & mean - std    \\
Surface geopotential height (z)  & 0                                                          & CERRA   & m            & Forcing    & max           \\
Forcing variables                &                                                            & Derived & Units        & Type       & Normalisation \\
Land-sea mask (lsm)              &                                                            & No      & -            & Forcing    & None          \\
Cosine of Julian day             &                                                            & Both    & -            & Forcing    & None          \\
Sine of Julian day               &                                                            & Both    & -            & Forcing    & None          \\
Cosine of local time             &                                                            & Both    & -            & Forcing    & None          \\
Sine of local time               &                                                            & Both    & -            & Forcing    & None          \\
Cosine of latitude               &                                                            & Both    & -            & Forcing    & None          \\
Sine of latitude                 &                                                            & Both    & -            & Forcing    & None          \\
Cosine of longitude              &                                                            & Both    & -            & Forcing    & None          \\
Sine of longitude                &                                                            & Both    & -            & Forcing    & None          \\
Cosine of the Solar zenith angle &                                                            & Both    & -            & Forcing    & None          
\end{tblr}
\end{table}

\newpage
\section{Technical definitions: CERRA climatology and verification metrics}
\label{app:metrics}

The aim of this study is to quantitatively compare forecasts from LAM and SGM against CERRA reanalysis, climatology and synoptic observations. In the following, the term ``observation'' will be used to represent an assumed ``truth'' which could be either the CERRA reanalysis or observations from a station. Technical definitions for the climatology and the verification metrics used in the study are provided below. 

\subsection{Climatology benchmarking using CERRA}

To assess forecast skill and to allow for the computation of anomaly-based scores, a CERRA climatology is computed. The climatology consists of the averaged values from 1985 to 2019 for each variable and relevant time of day -- specifically, 00, 06, 12, and 18 UTC -- on a per-day-of-the-year and per-grid-point basis. The year 1984 was excluded since the full year is not present in the CERRA dataset and 2020 was left out to keep the climatology independent of the test period (see subsection~\ref{subsec:training}).

\subsection{MSE, RMSE and RMSE skill score}

The forecast mean squared error (MSE) is defined as:
\begin{equation}
\text{MSE}_f= \frac{\sum_{i=1}^N(f_{i}-o_{i})^{2}}{N},
\end{equation}

with $f_{i}$ the forecast value of a model $f$, $o_{i}$ the corresponding observation, and $N$ the total number of data points.

The forecast root mean squared error (RMSE) is simply the square root of MSE:

\begin{equation}
\text{RMSE}_f= \sqrt{\text{MSE}_f}.
\end{equation}

To allow for the comparison of LAM and SGM RMSE values against CERRA, the RMSE skill score (RMSSS) is defined as follows: 

\begin{equation} \label{eq:rmsss}
\text{RMSSS} = 1 - \frac{\text{RMSE}_{\text{LAM}}}{\text{RMSE}_{\text{SGM}}}.
\end{equation}

To test the hypothesis that $\text{RMSE}_{\text{LAM}}$ and $\text{RMSE}_{\text{SGM}}$  metrics are equal, we define a confidence interval of the RMSSS value using a bootstrap procedure \cite{Price2024}. Specifically, the confidence interval was obtained via randomly resampling inferred fields for both numerator and denominator in Eq.~(\ref{eq:rmsss}), and re-estimate the RMSSS from the resampled data. By repeating  this process, we can then define a confidence interval for the RMSSS value, and determine a critical region for rejecting the null hypothesis. If the interval does not contain the null value, we reject the null hypothesis, indicating that $\text{RMSE}_{\text{LAM}}$ and $\text{RMSE}_{\text{SGM}}$ are significantly different.

\subsection{Mean Squared Skill Score against climatology}

To assess the skill of a forecast relative to climatology, the Mean Squared Skill Score (MSSS) with respect to climatology is computed:

\begin{equation}
 \text{MSSS} = 1 - \frac{\text{MSE}_{f}}{\text{MSE}_{c}},
\end{equation}

with $\text{MSE}_f$ and $\text{MSE}_{c}$ being the Mean Squared Error of the forecast model ($f$) being evaluated and the climatology ($c$), respectively.

\subsection{Forecast activity and normalised forecast activity}

The forecast activity (ACT) is defined as the standard deviation of the forecast anomaly with respect to the climatology:
\begin{equation}
\text{ACT}_f= \sqrt{\frac{1}{N}\sum_{i=1}^N \left[ (f_{i}-c_{i}) - \frac{1}{N}\sum_{j=1}^N(f_{j}-c_{j}) \right]^2}.
\end{equation}

The normalised forecast activity (NFA) is then the ratio of the activity of the forecast and that of the observations:
\begin{equation}
\text{NFA}= \frac{\text{ACT}_f}{\text{ACT}_o}.
\end{equation}

\subsection{Equitable Threat Score}
The Equitable Threat Score (ETS)\cite{nipen2024} is used to assess a model's ability to forecast extremes. For a specific  threshold, it measures how often forecasts and observations exceed or fall below this threshold. This allows us to categorize events into four types and count them as such: number of hits ($a$), number of false alarms ($b$), number of missed events ($c$), and number of correct rejections ($d$). The ETS for a specific threshold is then computed as follows: 

\begin{equation}
    \text{ETS} = \frac{a - \hat{a}}{a + b + c - \hat{a}},
\end{equation}

with $\hat{a}$  the number of accurate predictions of the event occurring by chance:

$$ \hat{a} = \frac{(a+b)(a+c)}{a + b + c + d}.$$

The ETS penalises forecasts with a high number of false alarms and event misses, with values ranging from -1/3 to 1. Negative values indicate that a random forecast would be more reliable than the actual forecast.

\end{appendices}

\end{document}